\documentclass[prb,amsfonts,amssymb,floats,twocolumn,superscriptaddress,aps,floatfix]{revtex4-2}
\renewcommand{\paragraph}[1]{\textit{#1.}\textemdash}
\newcommand{\ui}{\mathrm{i}}

\makeatletter
\renewcommand{\fnum@figure}{\hspace{11pt}FIG. \thefigure}
\makeatother

\usepackage{comment} 
\usepackage{graphicx}
\usepackage{pythontex}
\usepackage{bm}
\usepackage{amsmath}
\usepackage[utf8]{inputenc}
\usepackage{amssymb}
\usepackage{dsfont}
\usepackage{mathrsfs}
\usepackage[english]{babel} 
\usepackage{enumitem}
\usepackage{braket} 
\usepackage{bbold}  
\usepackage{listings}
\usepackage{booktabs} 
\usepackage[colorlinks=true, breaklinks=true, linkcolor=blue, citecolor=blue, urlcolor=blue]{hyperref} 
\usepackage{MnSymbol}
\usepackage{orcidlink}
\usepackage{float}

\makeatletter
\newcommand{\appendixtableofcontents}{%
  \section*{Appendix Contents}%
  \@starttoc{atoc}%
}

\newcommand{\appsection}[1]{%
  \section{#1}%
  \addcontentsline{atoc}{section}{\protect\numberline{\thesection}#1}%
}
\newcommand{\appsubsection}[1]{%
  \subsection{#1}%
  \addcontentsline{atoc}{subsection}{\protect\numberline{\thesubsection}#1}%
}
\newcommand{\appsubsubsection}[1]{%
  \subsubsection{#1}%
  \addcontentsline{atoc}{subsubsection}{\protect\numberline{\thesubsubsection}#1}%
}
\makeatother

\begin{document}

\title{Spectral Functions of an Extended Antiferromagnetic $S=1/2$ Heisenberg Model \\ on the Triangular Lattice}

\author{Markus Drescher \orcidlink{0000-0001-9231-1882}}
\thanks{markus.drescher@tum.de}
\affiliation{Department of Physics, Technische Universität München, 85748 Garching, Germany}%

 
\author{Laurens Vanderstraeten \orcidlink{0000-0002-3227-9822}}
\affiliation{Center for Nonlinear Phenomena and Complex Systems, Université Libre de Bruxelles, 1050 Brussels, Belgium}%

\author{Roderich Moessner}
\affiliation{Max-Planck-Institut für Physik komplexer Systeme, 01187 Dresden, Germany}%

\author{Frank Pollmann \orcidlink{0000-0003-0320-9304}}
\affiliation{Department of Physics, Technische Universität München, 85748 Garching, Germany}%
\affiliation{Munich Center for Quantum Science and Technology (MCQST), 80799 Munich, Germany}%

\date{\today}

\begin{abstract}
We study an extended spin-$1/2$ antiferromagnetic Heisenberg model on the triangular lattice, which includes both nearest- and next-nearest-neighbor interactions, as well as a scalar chiral term. This model exhibits a rich phase diagram featuring several competing phases: different quantum spin liquids and various magnetically ordered states, including coplanar $120^\circ$ order, stripe order, and non-coplanar tetrahedral order.
We employ large-scale matrix product state simulations optimized for GPUs to obtain high-resolution dynamical responses. Our calculations reveal the spectral features across both ordered and liquid regimes of the phase diagram, which we analyze in comparison with analytical predictions and field-theoretical approaches.
We identify unique signatures of the ordered phases in the form of gapless Goldstone modes at the ordering wave vectors.
Our results in the $J_1-J_2$ quantum spin-liquid regime are indicative of a $U(1)$ Dirac spin liquid.
In the chiral spin-liquid phase, we find signatures of spinons as the fractional excitations of the underlying theory, manifested as the onset of a two-spinon continuum that agrees with predictions from the Kalmeyer-Laughlin ansatz for the ground-state wave function, and collective modes that can be viewed as spinon bound states.
We discuss finite-size effects, their consistency with the presumptions from field-theory, and review the dynamical structure factor with regard to experimentally relevant features such as the occurrence of highly dispersive signals and the global distribution of spectral weight.
\end{abstract}

\maketitle

\section{Introduction}%

Frustrated magnetic spin systems exhibit very rich phase diagrams due to the intrinsically strong nature of their correlations~\cite{frustratedmagnetism}. The variety of states include in particular liquid phases and exotic ordering patterns.
Quantum spin liquids (QSLs) especially---highly entangled phases of matter characterized by a lack of magnetic order at zero temperature and the emergence of fractionalized quasiparticle excitations~\cite{Balents2010, Savary2016, Knolle2019}---have been the focus of intense research both in theory and experiment~\cite{Wen2002, Coldea2001, Broholm2020}.


\begin{figure}
    \centering
    \includegraphics[scale=1]{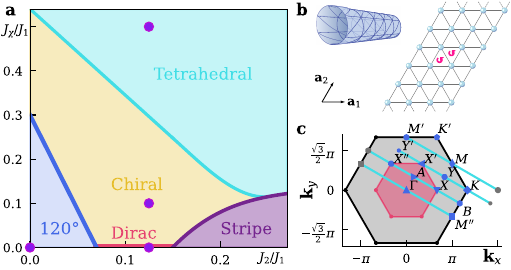}
    \caption{\textbf{a,} Schematic sketch of the phase diagram for the extended $J_1-J_2$ Heisenberg model with a chiral coupling $J_\chi$ on the triangular lattice (adapted from Ref.~\cite{Gong2017, Wietek2017}). The purple circles indicate points in parameter space where we computed the spectral function. \textbf{b,} Illustration of the geometry of the simulated system. ${\bf a}_{1/2}$ denote the Bravais lattice vectors. The simulations were performed for a geometry $\mathrm{YC}L$ with $L=6$, where the two dimensional lattice is mapped onto a cylinder and the periodic boundary conditions are recovered by translations of $L\cdot {\bf a}_2$. \textbf{c,} The Brillouin zone and the momentum cuts available for the applied cylinder geometry. Specific momentum points are highlighted. The violet shape indicates the reduced Brillouin zone of the spinons in the chiral spin liquid phase (cf. Ref.~\cite{Kalmeyer1989}).}
    \label{fig:phase_diagram}
\end{figure}

Following Anderson's seminal proposal of the resonating valence-bond (RVB) state~\cite{Anderson1973, Fazekas1974}, the quest for such novel phases of matter originated from the theoretical investigation of the nearest-neighbor Heisenberg model on the triangular lattice, which, however, evaded the expected ground state behavior. Instead, the lowest-energy configuration forms a coplanar three-sublattice $120^\circ$ spin order~\cite{HuseElser1988, Jolicoeur1990, Bernu1992, Bernu1994, Capriotti1999, White2007, Chernyshev2009}.
By contrast, the quantum dimer model on the triangular lattice does support a topologically ordered RVB state~\cite{Moessner2001}. Moreover, frustration in Heisenberg systems can be enhanced by introducing next-nearest-neighbor exchange interactions, which have been shown to drive a disordered ground state at intermediate coupling strengths~\cite{Manuel1999, LiBishopCampbell2015, BishopLi2016, Zhu2015, Hu2015, Iqbal2016}.
Its exact nature is still under debate with recent indications for an extended Dirac spin liquid regime~\cite{Hu2019, Sherman2023, Drescher2023, Wietek2024}.

Extending the $J_1-J_2$ Heisenberg model on the triangular lattice by a scalar chiral term leads to the emergence of a chiral spin-liquid phase~\cite{Gong2017, Wietek2017}. The proposal of such a chiral spin liquid as a novel disordered phase of matter originated from intensified theoretical investigations in the context of Anderson's proposition~\cite{Anderson1987, AndersonScience1987}:
Thanks to the pivotal contributions by Kalmeyer and Laughlin~\cite{Kalmeyer1987, Kalmeyer1989} who related Laughlin's original proposal of a wave function for the fractional quantum Hall effect (FQHE)~\cite{TsuiStormerGossard1982, Laughlin1983} to the RVB state~\cite{Kalmeyer1987} and finally to the ground state of a two-dimensional spin-$1/2$ Heisenberg antiferromagnet~\cite{Kalmeyer1989}, the advanced analytic tools opened the way towards new exotic phases: Wen, Wilczek and Zee's resulting conjecture of a chiral spin liquid~\cite{Wen1989} initiated an ongoing investigation of lattice models that stabilize chiral phases due to quantum fluctuations for various geometries such as the square and Kagome lattices~\cite{Bauer2014, He2014, ZhuGongSheng2015, GongZhuBalentsSheng2015, Kumar2015, Wietek2015, Haghshenas2019, Niu2022, Sun2024}. In particular, the extended Heisenberg antiferromagnet on the triangular lattice with a scalar chiral interaction that breaks time-reversal symmetry results in an extended phase diagram exhibiting a gapped chiral spin liquid and a spontaneously symmetry-breaking order where the four spins in a unit cell point to the corners of a tetrahedron. Inspired by previous numerical works on the subject~\cite{Gong2017, Wietek2017}, we sketch a schematic phase diagram in Fig.~\ref{fig:phase_diagram} that can serve as a guide through the subsequent parts of this work.

Despite the decades-long theoretical exploration of frustrated spin systems, the limitations in analytical, semiclassical and numerical methods still pose significant challenges to the understanding of the underlying physical processes on non-bipartite lattices as the triangular arrangement~\cite{Starykh2006, Chernyshev2006, Chernyshev2009, Zheng2006, Zheng2006PRL, Ghioldi2018, Ferrari2019}, especially with regard to the excitation spectrum~\cite{Gohlke2017, Verresen2018}.
However, recent progress in experiments and material composition~\cite{Bag2024, scheie_battista_2024, Cao2025} has made the realization of a spin-liquid phase a relevant option that requires a toolbox at hand to identify these uniquely in experimental probes and potentially resolve long-standing controversies about their nature.
Alongside the experimental developments---promising candidate materials are particularly rare-earth materials such as $\mathrm{Yb}$-based delafossites~\cite{Ranjith2019, Ranjith2019Dec, Zhang2021, Dai2021, Scheie2024}  or the compound $\mathrm{YbZn_2GaO_5}$~\cite{Bag2024}---, the latest progress in the application of tensor network methods to the dynamics of two-dimensional spin systems~\cite{Verresen2019, Vanderstraeten2019, VanDamme2021, Sherman2023, Drescher2023} provides access to spectral data in improved resolution and quality. In addition to the matrix-product state (MPS) formalism~\cite{Schollwoeck2011, Hauschild2018, Vanderstraeten2019}, numerous works on the chiral spin-liquid phases using projected entangled pair states (PEPS)~\cite{Poiblanc2017, Chen2018, Niu2022}, as a method intrinsically designed for two-dimensional systems, have resulted in proving the applicability of such methods as well to chirality-breaking models~\cite{HasikVanderstraeten2022, HasikCorboz2024}. The numerical advances have thus led to numerous works specifically on the chiral phases in the Hubbard and the descendant Heisenberg models establishing the existence of chirality-breaking ground state physics~\cite{GongZhuSheng2014, Szasz2020, Szasz2021, Kuhlenkamp2024, Budaraju2024}.

\begin{figure}
    \centering
    \includegraphics[scale=1]{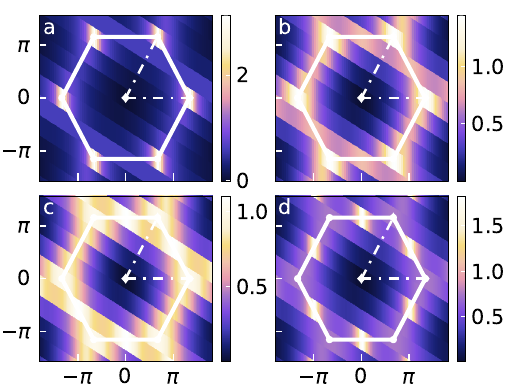}
    \caption{The static structure factor as defined in Eq.~\eqref{equ:ssf} for various phases: \textbf{a}, the $120^\circ$ ordered phase at $J_2=0$, \textbf{b}, the quantum spin liquid phase at $J_2/J_1 = 1/8$, \textbf{c}, the chiral spin liquid phase at $J_2/J_1 = 1/8$ and $J_\chi/J_1=0.1$ and \textbf{d}, the tetrahedrally ordered phase at $J_\chi/J_1=0.5$ ($J_2$ as in \textbf{c}). We observe clear signatures of Goldstone modes in the ordered phases, i.e., at the $K$-points for the $120^\circ$ order and at the $M$-points for the tetrahedral order. In the liquid phases, the distribution of spectral weight is more uniform and does not exhibit Bragg peaks as there is no ordering wave vector. The spectral weight at the $\Gamma$-point vanishes for all phases due to the spin-rotation symmetry of the model.}
    \label{fig:ssf}
\end{figure}

In this work, we aim to shed new light on the nature of the distinct liquid and ordered phases in the extended Heisenberg model on the triangular lattice as presented in Fig.~\ref{fig:phase_diagram} by computing spectral functions from large-scale state-of-the-art MPS simulations~\cite{WhiteAffleck2008, Gohlke2017, Verresen2018, Verresen2019}.
Applying graphics processing unit (GPU) infrastructure for our dynamical simulations enabled us to reach substantially larger virtual bond dimensions of up to $\chi = 4000$ compared to previous works.
The article is organized as follows: In Sections~\ref{sec:model} and \ref{sec:Numerical Methods} we introduce the model Hamiltonian and discuss in more technical detail the numerical protocols we used to obtain our results. Sections~\ref{sec:120} to \ref{sec:tetra}  compare the field-theoretical expectations and the numerical findings for the individual phases under consideration, i.e., the $120^\circ$-ordered phase, the putative Dirac spin liquid (as which we will refer to this phase hereafter), the chiral spin-liquid phase and the tetrahedrally ordered phase.
Our high resolution data provides a clear  distinction between the ordered and liquid phases: While the ground state order from spontaneous symmetry breaking results in Goldstone modes with a high concentration of spectral weight at the corresponding ordering momenta, the liquid phases come with a broad distribution of spectral intensity in the dynamical spin structure factor. Both the $120^\circ$ phase and the tetrahedrally ordered region exhibit avoided quasiparticle decay of magnon branches in specific areas of the Brillouin zone. The candidate Dirac and chiral spin liquids, contrarily, are characterized by pronounced low-energy continua in their spectra whose faithful resolution seems unique to the technique based on dynamical correlations. Regarding the shape of the continua in the chiral spin-liquid phase, we identify therein subtle signatures of the Kalmeyer-Laughlin wavefunction~\cite{Kalmeyer1987, Kalmeyer1989} that manifest themselves in the energetic minima of the onset of the spinon continuum.
Across the different phases, we observe changes of the detectable energy gap that is comprised of a finite-size gap and potential phase-related physical gaps that are consistent with the theoretical expectations: While, coming from the $120^\circ$ order, the original finite-size gap at momentum point $K$ opens when passing through the Dirac spin liquid towards the chiral phases, the gap at $M$ is softest in the Dirac regime and the tetrahedral order where gapless excitations are expected to reside there. In section \ref{sec:conclusion} we conclude with a brief summary and outlook.


\section{Model}
\label{sec:model}
We consider an extended Heisenberg model with nearest and next-nearest-neighbor exchange interactions and a scalar chiral spin interaction term~\cite{Wen1989, Baskaran1989} on the triangular lattice, which is described by the following Hamiltonian~\cite{Bauer2014, Wietek2015, Wietek2017, Gong2017}:

\begin{align}
H &= J_1 \sum_{\langle i,j\rangle} \hat {\bf S}_i \cdot \hat {\bf S}_j
+ J_2 \sum_{\llangle i,j\rrangle} \hat {\bf S}_i \cdot \hat {\bf S}_j \nonumber \\
& + J_{\chi} \sum_{i,j,k \in \triangle, \triangledown} \hat {\bf S}_i \cdot \left(\hat {\bf S}_j \times \hat {\bf S}_k\right)\,.
\label{equ:Hamiltonian}
\end{align}

The sum in the scalar chirality term goes over all triangles of the lattice under which the chirality of the order of the spins is kept identical (as indicated in Fig.~\ref{fig:phase_diagram}\textbf{\color{blue}b}).
This term breaks time-reversal and parity symmetry, but preserves $SU(2)$. It leads to the emergence of multiple additional phases as shown in previous numerical investigations, among which there is a presumptive chiral spin liquid and for larger coupling constant $J_\chi$ a tetrahedrally ordered phase~\cite{Wietek2017, Gong2017} (cf. Fig.~\ref{fig:phase_diagram} for a schematic phase diagram).

When tuning the next-nearest neighbor coupling parameter $J_2$ at vanishing $J_\chi$, the system transitions from a coplanar order with spin orientations at a relative angle of $120^\circ$ respectively~\cite{HuseElser1988, Bernu1992, Capriotti1999, Chernyshev2006, White2007, Chernyshev2009} with increasing $J_2$ through a disordered phase potentially described by a Dirac spin liquid~\cite{Iqbal2016, Hu2019, Ferrari2019, Sherman2023, Drescher2023, Wietek2024} in the interval $0.07 \lesssim J_2/J_1 \lesssim 0.15$~\cite{Manuel1999, Kaneko2014, LiBishopCampbell2015, Zhu2015, Hu2015, BishopLi2016, Iqbal2016} around the classical phase transition point at $J_2/J_1 = 1/8$~\cite{Jolicoeur1990, Chubukov1992} to a coplanar stripe order~\cite{Chernyshev2009, Hu2015, Zhu2015}. The scalar chiral term finally leads to the emergence of an extended region of a chiral spin-liquid phase and, for sufficiently large coupling constants $J_\chi$, to an ordered state with a tetrahedral spin pattern~\cite{Wietek2017, Gong2017}.

\section{Numerical Methods}%
\label{sec:Numerical Methods}

We obtain the spectral function from the dynamical spin-spin correlations of the system:

\begin{equation}
S^{+-}({\bf k}, \omega) = \hspace{-0.3em}\int \hspace{-0.3em} \mathrm{d}t  \sum_{j} e^{\ui\omega t - \ui{\bf k} \cdot ({\bf r}_j - {\bf r}_{j_c})} \braket{{\hat S}^{+}_j(t){\hat S}^{-}_{j_c}(0)}\,.
\label{equ:dsf}
\end{equation}

For this purpose, we compute the ground state of the Hamiltonian using the density matrix renormalization group algorithm for infinite systems (iDMRG)~\cite{White1992, White1993, McCulloch2008} on a cylinder~\cite{Stoudenmire2012}. This means that there is a periodic closure of the boundary conditions along the circumference of the cylinder with size $L$ and infinite boundary conditions along the cylinder axis with size $L_x$.
The ground state correlations themselves allow us to compute the static spin-structure factor that discloses information about the order of the ground state:
\begin{equation}
    \chi({\bf k}) = \frac{1}{N} \sum_j e^{-\ui {\bf k} \cdot ({\bf r}_j - {\bf r}_{j_c})}\braket{{\hat S}^{+}_j{\hat S}^{-}_{j_c}}\,.
    \label{equ:ssf}
\end{equation}
Here, we already made use of the translation invariance of the ground state for the formulation given.
From the infinite matrix-product state (iMPS), we can construct a large ground state system by sequentially repeating the iMPS unit cell along the x-direction. This allows us to perform large-scale time evolutions without the perturbation signal touching the boundaries in the course of the simulation, which would falsify the result by producing artifacts that are not relevant for bulk physics and related experiments.

Applying the WII time-evolution algorithm~\cite{Zaletel2015, Paeckel2019} based on the matrix-product operator (MPO) formulation using $U(1)$ charge conservation~\cite{SinghPfeifferVidal2010, SinghPfeifferVidal2011, Hauschild2018}, we can perform discrete time-evolution steps on the ground state perturbed by a single-site spin operator such as $S^-_{j_c}$. The site of the initial perturbation $j_c$ is chosen in the center of the system such that smooth and undisrupted evolution is possible avoiding the aforementioned boundary artifacts.
The algorithm relies on an approximation of the time-evolution operator $U(\delta t)$ formulated in terms of an MPO that can be applied variationally to the MPS representing the quantum-mechanical wave function $\ket{\psi}$. The variational application can be based on singular-value decompositions (SVDs) or QR-decompositions~\cite{Schollwoeck2011}, where the latter is particularly advantageous when the simulation is performed on a GPU (see also Appendix~\ref{App:GPU} for further details).
Having computed the dynamical correlations, we can obtain the spectral signal by performing a discrete Fourier transformation in space, followed by a second one in time to obtain the frequency resolution. It is common practice to extend the simulated time series by linear prediction and apply the Fourier transformation in time to a convolution of the extended signal with a Gaussian window function in order to prevent strong Gibbs oscillations that would result from the transformation of finite-time data (see Appendix~\ref{App:Protocol} for further details). The Gaussian broadening is given as $\sigma = \sqrt{t_{\mathrm{sim}}^2/(-2\cdot \ln{\alpha})}$ where $t_{\mathrm{sim}}$ denotes the physical simulation time and $\alpha$ is an adjustable parameter to control the width of the Gaussian.
Note that the choice of the Gaussian broadening as applied before transforming the time series data to frequencies plays a crucial role in regulating the final spectral function. For the data presented in this work, we made a conversative choice ($\alpha = 10^{-11}$) for panels exposing a summary of the results along different momentum lines while for selected cuts, the faithfulness of the simulation (and corresponding suppression of artifacts) allows for a smaller Gaussian broadening ($\alpha = 10^{-6}$) as shown in the insets.
An illustration of the convergence supplied with concrete data plots can be found in the Appendix~\ref{sec:convergence_supp}.
Moreover, we can apply an interpolation scheme in momentum space before the transformation towards frequency space which procedure results in a higher momentum resolution (cf. Appendix~\ref{App:Protocol} for details on the applicability and implementation of this scheme).

In addition to the perturbation with a single-site operator as discussed above, we can also apply a combination of operators that has a well-defined $k_y$-momentum~\cite{Vanderstraeten2019, Drescher2023}:

\begin{equation}
    \hat S_{n_1}^-(k_y) = \frac{1}{\sqrt{L}} \sum_{n_2 = 0}^{L-1} e^{\ui n_2 k_y} \hat S_{n_1 {\bf a}_1 + n_2 {\bf a}_2}^-\,.
\end{equation}

Here, ${\bf a}_1 = (0, 1)^T$ and ${\bf a}_2 = (1/2, \sqrt{3}/2)^T$ denote the Bravais lattice vectors of the triangular geometry under consideration (cf. Fig.~\ref{fig:phase_diagram}\textbf{\color{blue}b}). The selection of $k_y$ fixes effectively the available momentum space of the simulation to the corresponding cut in Fig.~\ref{fig:phase_diagram}\textbf{\color{blue}c}. Choosing $k_y = 2\pi \cdot m/L$ would restrict us to the $m$th branch---the one for $m=0$ intersects with the origin of the Brillouin zone. This allows us to obtain for the same virtual bond dimension $\chi$ of the MPS dynamical correlations that are better converged and more accurate than for the single-site operator. Eventually, the separation of the individual momentum cuts also leads to distinct group velocities that determine the outer boundaries of the cone within which information spreading happens during the time evolution. Hence, it allows for individually optimized simulation settings (see Appendix~\ref{App:Overview} for details).

Using the speedup from nodes of four combined NVIDIA A100-GPUs (80GB RAM each)~\cite{li_numerical_2020, Pan2022, unfried_fast_2023}, we simulate the dynamical evolution of the spins for system sizes between $L_x = 51$ and $L_x = 126$ rings for the $\mathrm{YC}6$-cylinder~\cite{Zhu2015, Hu2019} and MPS bond dimensions of up to $\chi = 4000$ states. Our implementation conserves a global abelian $U(1)$ symmetry corresponding to $S^z$ conservation. The simulations in the $120^\circ$-ordered and the putative Dirac spin liquid phase have been performed at $\chi = 4000$ and a system size of $L\times L_x = 6 \times 51$ (the corresponding MPO bond dimensions are $\chi_{\mathrm{MPO}} = 23$ and $\chi_{\mathrm{MPO}} = 38$). With the chiral interaction present, the maximum MPO virtual bond dimension increases to $\chi_{\mathrm{MPO}} = 80$, restricting the feasible MPS bond dimension to $\chi = 2000$ in our simulations at a system extension of $L_x = 51$ and $L_x=81$. In the $120^\circ$ phase, momentum-resolved time evolutions have been performed for $L_x~=~126$ and $\chi = 2000$, in the chiral phases we additionally computed $k_y$-resolved correlations for $L_x = 51$ with the same bond dimension. All simulations have been performed with a single time-step size of $\delta_t = 0.04 \,J_1$. The data presented in the main text was obtained exclusively for the cylinder geometry $\mathrm{YC}6$, where the periodicity is closed by a $6$-fold translation by the lattice vector ${\bf a}_2$. Results for different geometries and larger circumferences than $L=6$---we could partially converge the simulations up to $L=12$---are discussed in Appendix~\ref{App:Overview}. We comment on finite-size effects and artifacts due to truncation errors.

We can complement these time-evolution simulations by a direct variational computation of the momentum-resolved excitation spectrum by means of the MPS quasiparticle ansatz, an MPS extension of the single-mode approximation~\cite{haegeman_variational_2012, Vanderstraeten2019, VanDamme2021}. This variational approach yields the lowest-energy states regardless of its spectral weight, and therefore gives direct access to low-lying modes that only give faint signals in the spectral function.

\section{$120^\circ$ Ordered Phase}
\label{sec:120}

At the Heisenberg point (both $J_2=0$ and $J_\chi=0$)---Anderson's originally proposed candidate for a novel disordered state~\cite{Anderson1973, Fazekas1974}---, the ground state of the system exhibits coplanar $120^\circ$ Néel order~\cite{Jolicoeur1990, Bernu1992, Capriotti1999, White2007, Chernyshev2009}. Such an ordered state that breaks the continuous spin rotation symmetry of the Hamiltonian yields gapless Goldstone modes as part of the low-energy excitation spectrum. The corresponding ordering vectors reside at the corners of the Brillouin zone (the $K$ points, cf. Fig.~\ref{fig:phase_diagram}\textbf{\color{blue}c}). In a combined effort, the three-sublattice coplanar $120^\circ$ order could be established by complementing results from semiclassical analytical techniques such as higher-order spin-wave theory~\cite{Chubukov1994, Chernyshev2006, Starykh2006, Chernyshev2009} and series expansion~\cite{Zheng2006, Zheng2006PRL} and numerical evidence~\cite{Bernu1992, Bernu1994, Capriotti1999, White2007}. The excitation spectrum displays already in the $1/S$ spin-wave expansion and in the series expansion strong renormalizations of the collective magnon mode compared to the first-order linear spin-wave theory (LSWT)~\cite{Starykh2006, Chernyshev2006, Zheng2006, Zheng2006PRL}. These include in particular rotonlike minima at the centers of the edges of the Brillouin zone (the $M$ points, see Fig.~\ref{fig:phase_diagram}\textbf{\color{blue}c}).

\begin{figure}[H]
    \centering
    \includegraphics[scale=1]{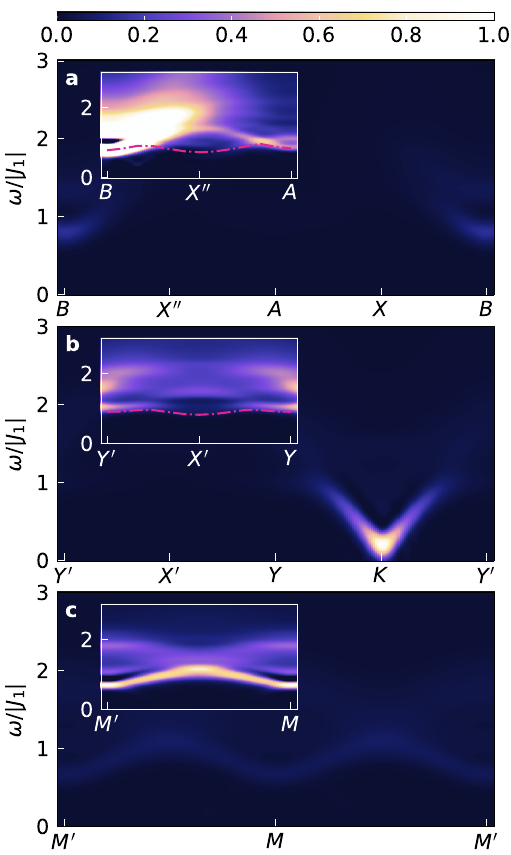}
    \caption{The spectral function at the Heisenberg point $J_2/J_1=0$ in the ordered phase with a coplanar spin configuration at a relative angle of $120^\circ$. The bond dimension used for the simulation is $\chi = 4000$. The geometry is $\mathrm{YC}6$ with cylinder length of $L_x = 51$. We simulate up to the total time of $t_{\mathrm{sim}} = 60\,J_1$ at a time step size of $\delta t = 0.04\,J_1$. The panels show the spectral function at various cuts accessible in the chosen geometry (see Fig.~\ref{fig:phase_diagram}\textbf{\color{blue}c}). The spectral weight as indicated in the color bar is normalized to the maximum of all displayed momenta. The largest spectral weight resides---as expected from the static structure factor in Fig.~\ref{fig:ssf}\textbf{\color{blue}a}---at the Bragg peak at momentum point $K$. The inset in \textbf{a} shows half of the cuts where, for the sake of better visibility of detailed features, an individual cutoff of the spectral weight has been introduced at $0.1\,S_{\mathrm{max}}$, i.e., ten percent of the overall maximum in the depicted momentum line. The insets in subplots \textbf{b} and \textbf{c} show data obtained from $k_y$-resolved simulations with $L_x=126$ at bond dimension $\chi = 2000$. The color bar is normalized to the maximum of the corresponding section. We use an interpolation scheme of the time series in momentum space to increase the resolution (cf. Appendix~\ref{App:Protocol}). The red dot-dashed lines in selected cuts indicate the lowest excitation mode as obtained from the MPS quasiparticle ansatz.}
    \label{fig:120}
\end{figure}

Figure~\ref{fig:ssf}\textbf{\color{blue}a} shows the static structure factor $\chi({\bf k})$ obtained from the ground state on the $6$-leg cylinder. We see pronounced Bragg peaks at the corners of the Brillouin zone, thus confirming the magnetic ordering of the ground state. Note that at the gapless points of the Goldstone modes at $K$ and $K^\prime$, the spectral weight is significantly higher than the peaks in the spectral weight distribution for the other phases considered. Figure~\ref{fig:120} shows the dynamical structure factor for only nearest-neighbor interactions. The main panels \textbf{\color{blue}a}-\textbf{\color{blue}c} show the full spectral function along the corresponding momentum cuts through the Brillouin zone. The explicit location of the cuts as given in the $\mathrm{YC}6$-geometry are drawn in Fig.~\ref{fig:phase_diagram}\textbf{\color{blue}c}. The color bar has \hfill been \hfill normalized\hfill to \hfill the\hfill maximum\hfill in

\begin{figure}
    \centering
    \includegraphics[scale=1]{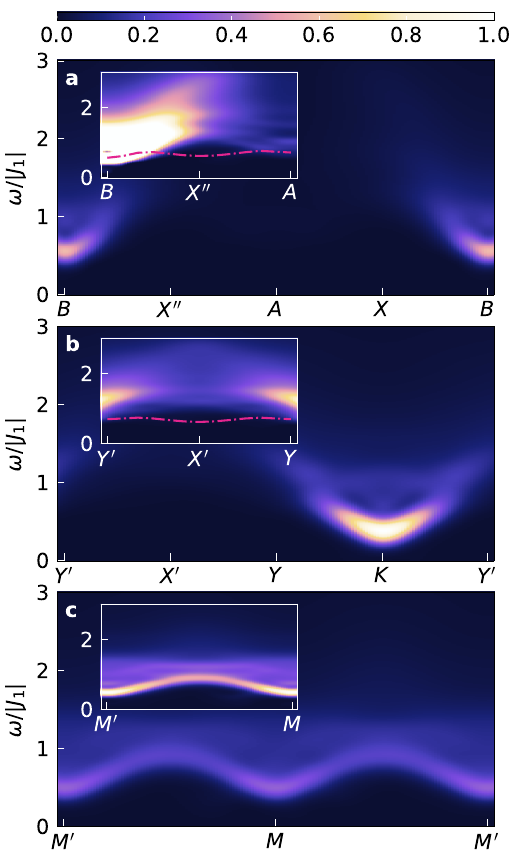}
    \caption{The spectral function for $J_2/J_1 = 0.125$ in the putative Dirac spin-liquid phase. Except for the coupling parameters, the simulation parameters are identical to the data presented in Fig.~\ref{fig:120}. The spectral weight in the inset has been normalized to the maximum of the respective cut (\textbf{b} and \textbf{c}) or to an individual cutoff of $0.1\,S_{\mathrm{max}}$ (subfigure \textbf{a}) aiming to improve the clarity of low-intensity features. The red dot-dashed line represents the result from the quasiparticle ansatz for the lowest-lying excitation.
    We observe distinct continua throughout the Brillouin zone.
    The most prominent low-energy feature is the minimum in the dispersion at $K$. There are rotonlike minima at the $M$ points. For a $U(1)$ Dirac spin liquid, they are expected to become gapless spinon bilinear excitations in the thermodynamic limit.}
    \label{fig:dsl}
\end{figure}

\noindent spectral weight over all accessible momenta. As readily perceivable from the analysis of the static structure factor---the integral of the dynamic structure factor over the frequencies for a fixed momentum yields the spectral weight of the static structure factor at this momentum point according to the sum rules---, the high concentration of spectral weight $S({\bf k}, \omega)$ at $K$ where the Goldstone mode comes down to form a gapless excitation dominates the overall spectral function. Note that in our numerical simulations, there will always be rather prevalent finite-size effects which occur due to the limitations in the circumference of the cylinder and the overall restricted number of spin sites we can take into account. Hence, the Goldstone mode in our spectral function at $K$ displays a minimum, but does not fully become gapless as expected in the thermodynamic limit. For sake of visibility, we plot details of the spectral signals in the insets for each momentum cut. In these insets, the scale of the color bar has been normalized to the maximum weight as it occurs in the depicted line in momentum space. Subplot \textbf{\color{blue}c} goes through the $M$-points and shows prominently the strong downward renormalization of the single-magnon branch at these center points of the edges of the Brillouin zone where the mode forms a rotonlike minimum as reported previously~\cite{Starykh2006, Zheng2006, Chernyshev2006, Oitmaa2020, Drescher2023}. Overall, we find conclusive agreement with complementary theoretical and experimental results~\cite{Mourigal2013, Ghioldi2018, Ferrari2019, Ghioldi2022, Scheie2024}.

In contrast to the prevalent expectations, when the magnon branch enters the two-magnon continuum in Fig.~\ref{fig:120}\textbf{\color{blue}a}, it does not simply decay, but in fact the mode is at least partially repelled from the onset of the continuum. This phenomenon, termed avoided quasiparticle decay or level repulsion, has been first reported for a triangular lattice Heisenberg model with an easy-axis anisotropy that gaps out the Goldstone modes and makes the problem more tractable by numerical means~\cite{Verresen2019}. The same physical behaviour could be demonstrated for the isotropic Heisenberg model~\cite{Drescher2023}, constituting the previous conjecture of mode repulsion a genuine feature of the nearest-neighbor Heisenberg model on a frustrated geometry. We support the finding in our full spectral function by the lowest-excitation results obtained from the quasiparticle ansatz for MPS (denoted by the red dot-dashed line in Fig.~\ref{fig:120}\textbf{\color{blue}a}). Even though the spectral weight of the repelled mode is very low, the excitation ansatz captures the resulting additional rotonlike minimum at $X^\prime$ (inset of Fig.~\ref{fig:120}\textbf{\color{blue}a}) and gives good agreement with the spectral function obtained from dynamical simulations. The inset in Fig.~\ref{fig:120}\textbf{\color{blue}b} shows another local minimum of the lowest magnon branch at the related point $X^\prime$ that results again from avoided magnon decay and has been confirmed by the excitation ansatz dispersion. As $X$ or $X^{\prime\prime}$ and $X^\prime$ are not equivalent on the cylinder geometry applied in our simulations, we do not observe fully identical features at the distinct momentum points. The spectral weight at $X^\prime$ is clearer in its resolution even though it similarly becomes smaller as the repelled mode approaches its minimal energy. The phenomenon of level repulsion is caused by the strong spin interactions that dominate the spin-$1/2$ system with its immanent quantum fluctuations~\cite{Verresen2019}. As a result, we observe stable magnon modes throughout the Brillouin zone. The avoided quasiparticle decay manifested itself also in experimental realizations of triangular lattice Heisenberg antiferromagnets~\cite{Ito2017, Macdougal2020}.

Compared to previous data~\cite{Drescher2023}, we find a better resolution of the spectral function and a reduced magnitude of artifacts due to truncation for these simulations at bond dimensions $\chi = 4000$ for the full single-site operator evolution and $\chi=2000$ for the $k_y$-resolved simulation (with the latter performed at a system extension of $L_x = 126$).

\section{Candidate Dirac Spin Liquid}
\label{sec:DSL}

For intermediate next-nearest neighbor couplings $0.07 \lesssim J_2/J_1 \lesssim 0.15$ (with $J_\chi = 0$), a disordered phase is formed whose precise nature has been the subject of a long-lasting intense debate~\cite{Kaneko2014, Zhu2015, Hu2015, Saadatmand2016, Iqbal2016, ZhuMaksimovWhiteChernyshev2018, Hu2019}. While a number of works, in particular early DMRG results, suggest a gapped $\mathbb{Z}_2$ spin liquid~\cite{Zhu2015, Hu2015, Saadatmand2016, Jiang2023}, variational Monte Carlo simulations found the Dirac spin liquid ansatz wave function to yield the lowest variational energy~\cite{Kaneko2014, Iqbal2016}. Recent studies have reported evidence of gapless modes in the paramagnetic regime~\cite{Hu2019, Budaraju2024}, while also previous dynamical and variational Monte Carlo simulations found excitation signatures compatible with the picture of a Dirac spin liquid~\cite{Ferrari2019, Sherman2023, Drescher2023}. A comprehensive study of the disordered phase via analyzing the overlaps of states from extensive exact diagonalization and trial wave functions demonstrated the key role of quantum electrodynamics in 2 + 1 dimensions ($\mathrm{QED}_3$) as the fundamental principle that captures the physics of the triangular lattice Heisenberg antiferromagnet~\cite{Wietek2024}. $\mathrm{QED}_3$ plays the role of an emergent effective field theory in 2-dimensional frustrated magnetic systems that describes algebraic spin liquids, also known as Dirac spin liquids (DSL) as the $N_f = 4$ fermion flavors exhibit gapless Dirac nodes~\cite{Hastings2000, Wen2002, Hermele2004, Hermele2005}. There are two main sorts of low-energy excitations that determine the spectrum of spin-1 spectral functions as we can probe it: On the one hand, pairs of spinons that are introduced in the parton construction as fermionic operators~\cite{Wen2002} can form bilinear excitations that lead to gapless cone-shaped continua in the thermodynamic limit at the $M$ points~\cite{willsher_dynamics_2025}. On the other hand, the coupling of the fermions to a $U(1)$ gauge field in the fundamental field theory introduces the possibility of relevant gauge fluctuations within the parton approach of the DSL. One important class of excitations are the magnetic monopoles---instanton events that insert a global $2\pi$ flux in the gauge background---whose quantum numbers on non-bipartite lattices have been determined in Refs.~\cite{Song2019, Song2020}. It turns out that the relevant triplet monopoles as detectable in our protocol reside at the corners of the Brilluoin zone. Again, as in the case of the spinon bilinear excitations, the monopoles should result in conic continua featuring a gapless spectrum at the $K$ points~\cite{willsher_dynamics_2025}. Investigations on the finite-size scaling of the monopole excitations on the triangular lattice using the construction of explicit variational wave functions alongside a subsequent Gutzwiller projection in order to maintain the physical spin Hilbert space indicate that both triplet and singlet monopoles become gapless in the thermodynamic limit~\cite{Budaraju2024}, thus supporting the conjecture of a gapless $U(1)$ Dirac spin liquid.

In the light of these field theoretical insights, we present spectral data at high resolution as a basis of comparison and a guide for inelastic neutron scattering experiments~\cite{Scheie2024, Bag2024}. In constrast to the $120^\circ$ Néel order, the paramagnetic phase comprises two distinct topological sectors on cylinders with an even number of legs. This can be attributed to the presence of spinon degrees of freedom~\cite{HeShengChen2014}.  Running iDMRG on the $\mathrm{YC}6$-cylinder yields a wave function representation in the even sector whose entanglement spectrum is symmetric around the total $S^z$ quantum number $S^z_{\mathrm{tot}} = 0$~\cite{Hu2015}. The odd sector---referring to the parity of the number of bonds cut by a line along the cylinder circumference in the resonating valence bond picture---can be reached either by an adiabatic flux insertion of $\theta = 2\pi$ or by leaving out a single site on each boundary of the cylinder during the first sweeps of iDMRG~\cite{Zhu2015, Hu2019}. For the data presented, we applied the second protocol. The resulting wave function features more isotropic correlations on the $L=6$-cylinder compared to the even sector, a slightly lower energy per site and an entanglement spectrum centered around $S^z_{\mathrm{tot}} = -0.5$, indicating the manisfestation of spinon particles at the edges of the cylinder.

We perform the time evolution for a system size with $L_x = 51$ at an MPS bond dimension of $\chi = 4000$. Compared to the previously discussed data for up to $\chi = 2000$~\cite{Drescher2023}, we observe a significant reduction of truncation effects such as negative spectral weight at long evolution times. Already the static structure factor (Fig.~\ref{fig:ssf}\textbf{\color{blue}b}) suggests with its broad distribution of spectral weight the absence of a distinct order. Regardless of the anisotropies due to the cylinder symmetry, we observe a maximum of spectral weight at the $K$ points with slightly weaker accumulations of weight at the $M$ points (which evens out for selected $M$ points for different geometries such as $\mathrm{XC}6$, see Appendix~\ref{App:Overview}). The most prominent feature we observe in the spectral function presented in Fig.~\ref{fig:dsl} is consequently a low-energy mode with an energetic minimum at $K$. There is a similar dispersive mode at the $M$ points (Fig.~\ref{fig:dsl}\textbf{\color{blue}c}). In both cases, we note a distinctive continuum that seamlessly emerges from the lowest mode in the spectrum. The pronounced two-spinon continuum is a key feature of the spectral function throughout the Brillouin zone. Note that in particular the gap between the onset of the continua and the mode closes with increasing bond dimension, i.e., accuracy, becoming basically gapless in the data presented in Fig.~\ref{fig:dsl} (see also the Appendix~\ref{App:Overview} for an explicit illustration of this aspect). As the Dirac cone of the spinons in the DSL mean-field theory occur at $\pm {\bf Q} = \pm (\pi/2, \pi/2\sqrt{3})$, we expect the spinon bilinear excitations to become gapless at the $M$ points. On the cylinder geometry, however, finite size effects inhibit this scenario. Instead even the spectrum within mean-field theory acquires a clear gap at $M$ and related points~\cite{Drescher2023}. Specifically the investigation and detection of gapless phases on cylinder geometries requires appropriate diligence~\cite{Ferrari2021}. Compared to the roton-minimum in the $120^\circ$ Néel order, however, the energy gap at $M$ softens when entering the liquid phase. This can be seen best by examining the spectral profiles given in Fig.~\ref{fig:spectral_cuts_KMX}\textbf{\color{blue}b}. We also observe there the shift of spectral weight from a massive concentration at the $K$ point for the three-sublattice order (Fig.~\ref{fig:spectral_cuts_KMX}\textbf{\color{blue}a}) towards a more balanced distribution in the DSL phase.
The origin of the distinct mode at $K$ is not fully understood yet. Interestingly, this minimum in the excitation spectrum occurs exactly at the momentum points where field-theoretical analysis locates the triplet monopole excitations as elucidated above~\cite{Song2019, Song2020}. On the finite-size cluster, the eigenstate at $K$ in the low-energy ED spectrum has significant overlap with the triplet monopole ansatz wave function~\cite{Wietek2024}. The excitation energy of this mode lies around $0.5\,J_1$ and is as such perfectly consistent with our findings. This value of the gap agrees with the results from explicit finite-size scaling in Ref.~\cite{Budaraju2024}. A complementary approach uses a self-consistent random phase approximation (RPA) for the $J_1-J_2$ Heisenberg model on the triangular lattice to obtain the dynamical structure factor for a $U(1)$ Dirac spin liquid~\cite{willsher_dynamics_2025}. This methods yields a branch exhibiting a rotonlike minimum at $K$ in remarkable agreement with the complete spectral function presented here. In the context of RPA, this single mode can be interpreted as a two-spinon bound state, termed a \textit{spinon exciton}~\cite{willsher_dynamics_2025}, that carries---in accordance with our numerical data---the highest spectral weight. At the transition to the $120^\circ$ order, this mode condenses at the designated wave vector of the minimum, thereupon forming the familiar spin order~\cite{Dupuis2019, willsher_dynamics_2025}. Hence, this mode can be regarded as a signature of the proximate Néel order.

Moreover, in Fig.~\ref{fig:dsl}\textbf{\color{blue}a}, we can identify resemblances to the previously discussed $120^\circ$ order. While the continuum in this case is broader and smoother than the two-magnon equivalent, we still detect weak modes below it which becomes especially clear from the quasiparticle ansatz (the red line in the inset). This is reminiscent of the level repulsion in the ordered case and could be caused by a similar mechanism for a two-spinon bound state entering the continuum.

Despite major finite-size effects, our results underline the relevance of the Dirac spin liquid for the intermediate paramagnetic phase. Moreover, they demonstrate the vital importance of gauge fluctuations in the DSL parton ansatz. Even though the technological advances will allow us to achieve enhanced precision, the available data for various cylinder geometries and circumferences (see Appendix~\ref{App:Overview}) indicate that finite-size effects in the accessible system sizes up to $L=12$ are still predominant as the physical processes are substantially influenced by the restrictions on the available momenta.

\section{Chiral Spin Liquid Phase}
\label{sec:CSL}

Starting from the Dirac spin-liquid phase, we can tune the additional chiral interaction to a finite coupling strength $J_\chi>0$ and transition to an extended regime where a chiral spin liquid (CSL) is stabilized~\cite{Wietek2017, Gong2017}.
The spin Hamiltonian in Eq.~\eqref{equ:Hamiltonian} explicitly breaks time-reversal and inversion symmetry~\cite{Wen1989, Baskaran1989}. The chiral spin-liquid wavefunction is of Kalmeyer-Laughlin type, i.e., it can be identified as the spin analogue of the $\nu = 1/2$ bosonic fractional quantum Hall state~\cite{Kalmeyer1987, Kalmeyer1989, Laughlin1990}. This state features a gapped bulk excitation spectrum along with gapless chiral edge modes~\cite{Kalmeyer1989, Wen1991}. The edge modes are described by an $SU(2)_1$ Wess-Zumino-Witten conformal field theory (CFT)~\cite{Wen1991}. On the infinite cylinder geometry, this edge spectrum can be detected from the entanglement spectrum at a bond between two rings, cutting the cylinder effectively into two parts~\cite{LiHaldane2008, QiKatsuraLudwig2012}.
Sorting the entanglement spectrum by momentum quantum numbers $k_y$ (which denote the transverse momentum along the cylinder circumference) yields a counting of energy levels that follows the CFT predictions~\cite{MooreHaldane1997, CincioVidal2013, Szasz2020}.
The existence of fractionalized excitations that carry spin-$1/2$ in the CSL introduces an additional method to probe such a phase: By adiabatically inserting a flux of $2\pi$ through the cylinder, we create a pair of spinons which are separated in the course of this pumping experiment to the opposite ends of the cylinder~\cite{Szasz2020, Kuhlenkamp2024}. This results in a net pumping of exactly spin $1/2$ across the cross section of the cylinder, indicating that the underlying CSL has a Chern number of $1/2$ which characterizes the state as the Kalmeyer-Laughlin wave function related to the $\nu = 1/2$ fractional quantum Hall effect~\cite{GongZhuSheng2014}.
We demonstrate these physical principles at the example of our iDMRG simulation in Appendix~\ref{App:GS}.

From the inspection of the static structure factor in Fig.~\ref{fig:ssf}\textbf{\color{blue}c}, we see that---similarly to the case of the Dirac spin liquid---there is a broad distribution of spectral weight in the Brillouin zone with even less pronounced maxima than for the DSL in Fig.~\ref{fig:ssf}\textbf{\color{blue}b}. We observe a shift of spectral weight from the $K$ points to the $M$ points, which demonstrates the influence of multiple adjacent phases such as the $120^\circ$ order---having Goldstone modes at the $K$ points---and the tetrahedral order exhibiting the gapless modes at all $M$ points. Due to the cylinder geometry and the breaking of the $C_6$-rotational symmetry, not all $M$ points, however, appear equivalent in this regime.

\begin{figure}[H]
    \centering
    \includegraphics[scale=1]{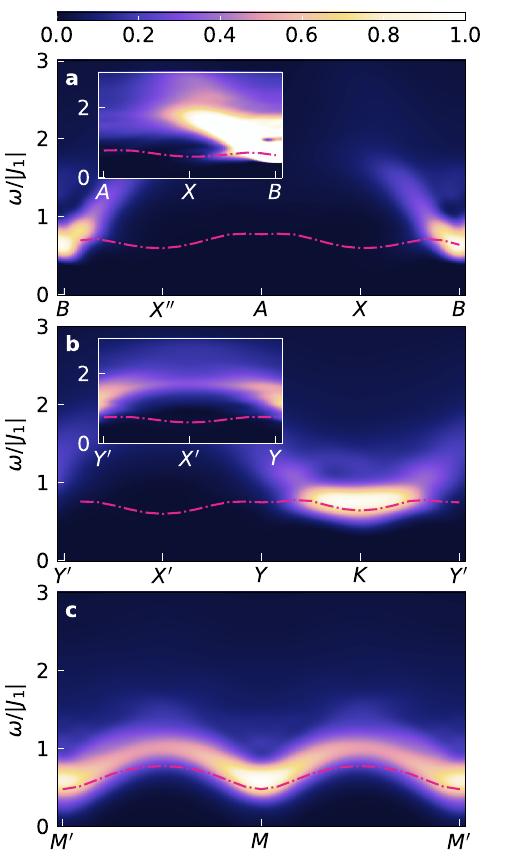}
    \caption{The spectral function in the chiral spin-liquid phase for $J_2/J_1 = 0.125$ and $J_\chi = 0.1$ at bond dimension $\chi = 2000$. The simulation was performed on a $\mathrm{YC}6$ cylinder with $L_x = 81$ (except for the inset in panel~\textbf{a} that was obtained form a momentum-resolved simulation with $L_x = 51$).
    The spectral weight in the inset in subplot \textbf{a} has been restricted to 10 percent of the maximum weight in the displayed momentum line. We observe a globally balanced distribution of the spectral intensity. The red dash-dotted line indicates the lowest-energy excitation as obtained from the variational quasiparticle ansatz. We observe a dispersive mode at the $B$ point in contrast to the almost flat feature at $K$. At the $K$ point, the $M$ points and---slightly visible in the inset in~\textbf{a}---at $X$, there is a broadening of the signal or a low-energy contribution of spectral weight that forms an energetic minimum at these specific points that can be attributed to spinon-contributions within the Kalmeyer-Laughlin ansatz~\cite{Kalmeyer1987} (see also Fig.~\ref{fig:spectral_cuts_KMX}, the main text and the discussion on convergence in Appendix~\ref{sec:convergence_supp}).}
    \label{fig:spectral_csl}
\end{figure}

Figure~\ref{fig:spectral_csl} shows the dynamical structure factor we obtained within the CSL phase at $J_\chi/J_1 = 0.1$ and $J_2/J_1 = 0.125$. The absence of strong peaks and the smooth distribution of spectral weight is evident also from the dynamical data: Instead of very pronounced signals at specific momentum points such as $K$ in the $120^\circ$-ordered phase, we have a balanced view of the spectral function in all cuts whose maximum intensities vary hardly among each other. As a consequence, this is a prominent feature to be applied in the evaluation of experimental measurement data as the overall distribution of spectral intensities allows us straightforwardly to tell apart disordered phases from those with an ordered ground state. Similarly to the previously discussed phases, we observe a local minimum in energy at the $M$ points. Even though the gap at $M^\prime$ does not decrease continuously as we tune up $J_\chi$ to enter the tetrahedral order eventually where the face-centered points in the Brillouin zone host Goldstone modes (cf. Fig.~\ref{fig:spectral_cuts_KMX}\textbf{\color{blue}b} and Appendix~\ref{App:Overview}), the minimum in the dispersion approaches (up to finite-size effects) a gapless point in the ordered phase at large $J_\chi$ with steep linear descent in contrast to the cosine-like dispersion of the potential spinon bound state in Fig.~\ref{fig:spectral_csl}\textbf{\color{blue}c}.
The $M^\prime$ point provides an illustration of the subtleties of interpreting finite-size effects in numerically accessible systems: Whereas the gap at $M$ in the Dirac spin liquid (supposedly originating predominantly from the small cylinder circumference) is expected to become a gapless continuum in the thermodynamic limit, the gap increases in the chiral spin-liquid domain as an intrinsically gapped phase of matter with the corresponding mode softening again under the transition into the tetrahedral order with its gapless Goldstone excitations.
These characteristics are clarified by spectral functions for two parameters sweeps in $J_\chi$ for $J_2 = 0$ and $J_2/J_1 = 0.125$ that are presented in Appendix~\ref{App:Overview}.
There is a well-developed continuum visible around the mode at $M$ in Fig.~\ref{fig:spectral_csl}\textbf{\color{blue}c}, in particular for energies above it. However, especially at $M^\prime$, the continuum seems to extend as well to lower energies (Fig.~\ref{fig:spectral_csl}\textbf{\color{blue}c} and Fig.~\ref{fig:spectral_cuts_KMX}\textbf{\color{blue}b} inset). This feature, as it is related to a full faithful representation of the continua in the model that can hardly be captured except for by an elaborate dynamical simulation, is not completely realized by the quasiparticle ansatz (the red dot-dashed line).
The variationally optimized single-mode ansatz attains between the $M$ points an approximation of the lower edge of the continuum. The factual distribution of the spectral weight above it shows a deviation from the lower onset of the continuum obtained from the quasiparticle ansatz, in contrast to the DSL phase.

The mode at the $K$ point satisfies the expectation for a gapped chiral phase: The potentially gapless mode in the DSL at $K$ (which naturally still reveals a finite-size gap in our simulations) gaps out with increasing values of $J_\chi$, hence leading to a gradual opening of the excitation gap at this momentum point. This is nicely illustrated in Fig.~\ref{fig:spectral_cuts_KMX}\textbf{\color{blue}a} and the corresponding inset. A similar mechanism might be at work at the $M$ points: Whereas in the extended DSL regime, the excitation spectrum there will develop a gapless continuum in the thermodynamic limit---effectively resulting in some low-energy signal with a finite-size gap---, the excitations at $M$ will open up a gap in the CSL phase, leading to an increased gap in our findings (Fig.~\ref{fig:spectral_cuts_KMX}\textbf{\color{blue}b}), before they become gapless Goldstone modes in the tetrahedrally ordered state (see Section~\ref{sec:tetra}).
The shape of the bound-state at $K$, however, changes notably under the transition 
from the\linebreak DSL\hfill to\hfill the\hfill CSL\hfill phase:\hfill The\hfill sharp\hfill rotonlike\hfill minimum

\noindent becomes an extended, rather flat dispersion with a broad maximum in the intensity and a minor dip at $K$ (which is readily substantiated by the quasiparticle ansatz calculation, see the red line in Fig.~\ref{fig:spectral_csl}\textbf{\color{blue}b}).
This mode can be related to the spin-$1$ collective excitation of the triangular lattice antiferromagnet that Kalmeyer and Laughlin obtained from a variational ansatz based on their wave function~\cite{Kalmeyer1989}. Their collective mode also shows a minimum at $K$ and a comparable energy gap at $M$.
Remarkably, the RPA diagrammatic computation finds a similar mode at $K$ in the chiral regime~\cite{willsher_dynamics_2025} as observed in our spectral function, with the distinction between Dirac spin liquid and chiral spin liquid being comparable: In the former case, the two-spinon bound state forms a distinct minimum in energy at $K$, while a similar mode in the chiral setting exhibits a flat feature in the dispersion. Nevertheless, in both cases, the field-theoretical approximation deviates from our results with regard to the continua: Figs.~\ref{fig:dsl}\textbf{\color{blue}b} and \ref{fig:spectral_csl}\textbf{\color{blue}b} present pronounced continua above and in proximity of the minima in the dispersion, which are not captured by the RPA-based approach. This circumstance hints at the importance of gauge fluctuations in the parton theory that go beyond the random-phase diagrammatics.

In contrast to the $K$ point, we observe at $B$ in Fig.~\ref{fig:spectral_csl}\textbf{\color{blue}a} a strong dispersive mode that has a high concentration of spectral weight. There is a stretched continuum (see inset), below which there are remnants of spectral weight that extend from the mode around $B$. The lower bound of the spectral signal can be measured by the quasiparticle dispersion. It shows a minor minimum at $X$ and accordingly at $X^\prime$ (inset of Fig.~\ref{fig:spectral_csl}\textbf{\color{blue}b}). Even though the low-energy spectral weight is close-to-vanishing around the $X$ points, a closer inspection reveals there is a noteworthy distinct plateau at lowest excitation energies in the spectral profile as depicted in Fig.~\ref{fig:spectral_cuts_KMX}\textbf{\color{blue}c} and the inset therein. This kink occurs exclusively in the CSL phase ($J_\chi/J_1 = 0.1$ and $J_\chi/J_1=0.2$ at $J_2/J_1 = 0.125$ being presented in the inset). Due to its faint spectral weight in absolute values, it will not be detectable in the current experimental setups, but it might reveal its importance in the context of the underlying field theory: 
Not only are the faint low-energy continua at the $X$, $K$, and $M$ points unique to the chiral spin liquid phase, they can also be attributed directly to the spinon dispersion as discussed by Kalmeyer and Laughlin in their comprehensive investigation~\cite{Kalmeyer1989}. The single-spinon dispersion resulting from the properties of the Kalmeyer-Laughlin wave function on the triangular lattice exhibits minima at the corners of the reduced spinon Brillouin zone (cf. Fig.~\ref{fig:phase_diagram}\textbf{\color{blue}c}). These momenta correspond to the $X$ points in the full Brillouin zone. Hence, a two-spinon continuum would necessarily acquire minima at two-fold combinations of these specific momentum points. There are four possibilities to which the various $X$-points can sum up: $\Gamma$, the origin of the Brillouin zone, the $M$-points in the centers of the edges of the Brillouin zone, the $X$-points themselves and the $K$-points at the corners of the reciprocal unit cell. The spectral weight at $\Gamma$ vanishes as the ground-state wave function is symmetric under $SU(2)$ spin rotations. At all other among these momentum points, we observe low-energy onsets of continua, i.e., more precisely a local minimum in the lower edge of the continuum.
While at the $X$ points, subtle low-energy plateaus in the spectral function are suggestive of these continua, a general broadening of the signal and kinks in the low-energy part of the spectral profiles are indications for them at the $K$ and $M$ points (cf. insets in Fig.~\ref{fig:spectral_cuts_KMX}). A more comprehensive discussion of the convergence of these features and their occurrence in long-evolution data is provided in Appendix~\ref{sec:convergence_supp}.
The onset of continua with minima at $M$, $K$ and $X$ are therefore in agreement with the previous theoretical derivations by Kalmeyer-Laughlin~\cite{Kalmeyer1989}.

\section{Tetrahedral Phase}
\label{sec:tetra}
\begin{figure}[h]
    \centering
    \includegraphics[scale=1]{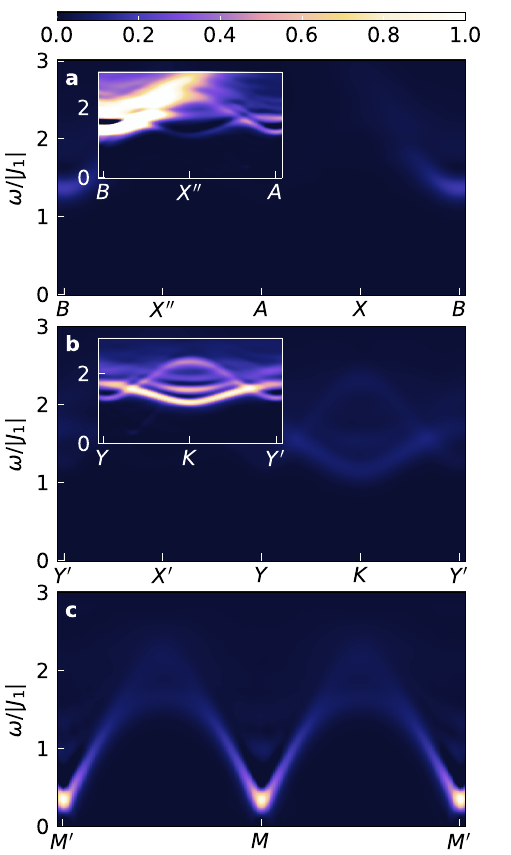}
    \caption{The spectral function in the tetrahedral phase for $J_2/J_1=0.125$ and $J_{\chi}=0.5$ at bond dimension $\chi = 2000$. This parameter regime comes with Goldstone modes symmetrically residing at the $M$-points of the Brillouin zone (cf. \textbf{c}). As in the simulation for the chiral spin-liquid state, we simulate up to $t_{\mathrm{sim}} = 60 \,J_1$ on a geometry $\mathrm{YC}6$ with $L_x = 81$. The inset in subfigure~\textbf{a} shows the phenomenon of level repulsion. For better visibility, the spectral weight has been cut at $0.1 \,S_{\mathrm{max}}$, where $S_{\mathrm{max}}$ denotes the maximum intensity of the spectral function in the chosen momentum line from $B$ to $A$. The inset in subfigure~\textbf{b} shows the dispersion of a single magnon branch as the lowest-energy mode. A discussion of the convergence of the repelled magnon branch is included in Appendix~\ref{sec:convergence_supp}.}
    \label{fig:spectral_tetrahedral}
\end{figure}

\begin{figure*}
    \centering
    \includegraphics[scale=1]{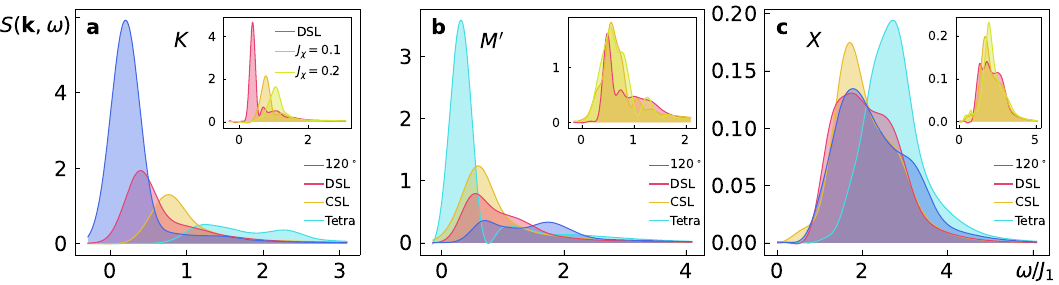}
    \caption{Comparison of the spectral function at high-symmetry momentum points throughout the four phases under consideration. The data in the $120^\circ$-ordered phase is taken at the Heisenberg point, i.e., $J_2=0$. The Dirac spin liquid (DSL) phase comes with the parameter $J_2/J_1 =0.125$ and the chiral spin liquid (CSL) is considered for $J_\chi/J_1=0.1$ and $J_2/J_1=0.125$. The tetrahedrally ordered phase (Tetra) has been evaluated at $J_\chi/J_1 = 0.5$ ($J_2/J_1=0.125$). The main plots show spectral data for a total simulation time of $36\,J_1$. The frequency was convoluted with a Gaussian broadening for parameter $\alpha = 10^{-11}$ to smoothen out Gibbs oscillations. The data for the $M$-point in the tetrahedral order has been obtained from a $k_y$-resolved simulation. So, too, were the data for $k_y = \pi/3$ in the $120^\circ$ phase in the inset in~\textbf{c} just like in the inset in Fig.~\ref{fig:120}\textbf{a}. We clearly notice the significant Bragg peaks at the condensation points of the Goldstone modes in the ordered phases ($K$ and $M$ respectively for the $120^\circ$ and tetrahedral phase). In the $120^\circ$ phase, the detectable gap at $K$ comes from finite-size effects. When tuning through the Dirac spin liquid into the chiral phases, we observe an opening of the gap. At $M^\prime$, on the contrary, the gap opens when transitioning from the DSL to the CSL and finally softens in the tetrahedral order. The insets contrast the Dirac spin liquid phase with the chiral spin liquid at two different values of $J_\chi$ ($J_2/J_1 = 0.125$ in all cases) after $60\,J_1$ and for a sharper broadening with $\alpha = 10^{-6}$. We note in the chiral spin liquid phase the onset of a potential two-spinon continuum at small energies at $X$, manifested through a plateau below the maximum, and at $K$ and $M^\prime$, indicated through broadened signals and low-energy kinks in the profiles.}
    \label{fig:spectral_cuts_KMX}
\end{figure*}

For sufficiently large coupling strength $J_\chi$, the chiral interaction term in Eq.~\eqref{equ:Hamiltonian} stabilizes a noncoplanar tetrahedral order across the whole range of next-nearest-neighbor exchange $J_2$ on the triangular lattice, where the spins in the four-site unit cell point towards the corners of a tetrahedron~\cite{KuboMomoi1997, Messio2011, Gong2017, Wietek2017}. With the DSL phase being stabilized around the classical phase transition at $J_2/J_1=0.125$, the CSL phase is located in this context in an extended region around the classical transition line between the coplanar $120^\circ$ order and the four-sublattice tetrahedral order~\cite{Messio2011, Gong2017}. The tetrahedral order survives in the quantum regime for sufficiently strong scalar chiral interactions~\cite{Gong2017, Wietek2017}. It is characterized by Bragg peaks at the $M$ points in the Brillouin zone related to the occurrence of gapless Goldstone modes at the corresponding ordering vectors~\cite{Messio2011}. The third ordered phase, the antiferromagnetic stripe order, is classically degenerate with the tetrahedral order for $J_2/J_1 > 0.125$ and $J_\chi = 0$. However, for finite $J_\chi > 0$, the tetrahedral state becomes immediately energetically favourable~\cite{Gong2017}. In the quantum spin model though, the stripe order occurs as the ground state in an extended region of the phase diagram (Fig.~\ref{fig:phase_diagram}).

We consider in our simulations the parameters $J_\chi/J_1 = 0.5$ and $J_2/J_1 = 0.125$. Figure~\ref{fig:ssf}\textbf{\color{blue}d} shows the static structure factor as obtained from the ground state correlations. It illustrates the underlying order via the distinct peaks at the $M$ points. The Goldstone modes emerging from the spontaneous symmetry breaking can be unambiguously identified in the spectral function in Fig.~\ref{fig:spectral_tetrahedral}\textbf{\color{blue}c}: The minimum in the dispersion is approached linearly by a magnon branch from both directions.
It comes along with a high concentration of spectral weight that is comparable to the $K$ points in the $120^\circ$ ordered phase.
In contrast to the previously discussed quantum spin-liquid phases, there are distinct sharp modes visible throughout the Brillouin zone. These are collective magnon excitations. Regardless of the finite-size effects, the gap at the $M$-point as pronouncedly present in the chiral spin liquid softens when transitioning towards the tetrahedral phase (see also Fig.~\ref{fig:spectral_tetrahedral}\textbf{\color{blue}b}). Note that the spectral intensity at $M$ is substantially higher than at all other accessible momentum points. This becomes apparent from the main panels in Fig.~\ref{fig:spectral_tetrahedral}.
The magnon mode forms a minimum at the $K$ point which coincides with minima at the other phases discussed in this work.
However, the gap increases (cf. also section~\ref{sec:CSL} and Fig.~\ref{fig:spectral_cuts_KMX}\textbf{\color{blue}a}) and the single modes are clearly distinguishable with a continuum setting in at higher energies.
The $K$ point shows thence a clear distinction between the gapped-out magnon branch here and the Goldstone mode in the $120^\circ$ order or the low-energy modes in the chiral and Dirac spin liquid phases.

Similarly to the previously discussed regimes, there is a dispersive signal around the $B$ point with an accumulation of spectral weight (Fig.~\ref{fig:spectral_tetrahedral}\textbf{\color{blue}a}). In analogy with the avoided quasiparticle decay observed for the coplanar order in section~\ref{sec:120}, the single-magnon branch entering the two-magnon continuum in the cut between $A$ and $B$ is repelled at the entrance point to the continuum and forms a stable collective mode below (see inset in Fig.~\ref{fig:spectral_tetrahedral}\textbf{\color{blue}a}). This feature at the $X$ point goes beyond spin-wave analysis. Due to sufficiently strong spin interactions, we observe a stable low-energy mode throughout the accessible momentum cuts in the Brillouin zone. This suggests that the phenomenon of avoided quasiparticle decay is a universal characteristics of Heisenberg models at sufficiently strong interactions~\cite{Verresen2019, Drescher2023}.

 


\section{Conclusion}
\label{sec:conclusion}

We have performed large-scale dynamical simulations on the triangular-lattice antiferromagnet and discussed the spectral data obtained at high resolution.
Not only does the numerics based on tensor-network formulations of two-dimensional quantum spin systems allow us to get increasingly precise answers about ground state properties and the corresponding excitation spectrum, the state-of-the-art application of GPU devices also opens up perspectives for experiments and the fundamental understanding of dynamical features required to identify previously ambiguous phases in frustrated magnetic systems.
In our work, we focused on four adjacent phases for the extended Heisenberg model on the triangular lattice: the coplanar $120^\circ$ order, the spin-liquid phases with the underlying field theories of the Dirac spin liquid and the chiral spin liquid, and the four-sublattice tetrahedral order.
Our findings are consistent with and indicative of microscopic properties derived from analytical considerations for the various phases, but go also beyond the realm of semiclassical analysis: Besides the expected gapless Goldstone excitations in the presence of spontaneous breaking of the spin-rotation symmetry, we find unique evidence of the phenomenon of avoided magnon decay both for the $120^\circ$ and the tetrahedral order. This result underlines the significance of spin interactions for the interpretation of Heisenberg systems on frustrated geometries. Level repulsion thus presents itself as a key signature of such models and provides a crucial guide for the interpretation of neutron scattering data for a diversity of compounds. The putative spin-liquid phases, on the other hand, are characterized by pronounced continua and distinct modes of highly dispersive quality such as at momentum point $B$ or more smooth behaviour as manifested prominently at the corners of the Brillouin zone. The continua especially are hard to obtain in a controlled way with any other method than the dynamical simulation of the full quantum system. With regard to this, our simulations have verified the importance of the spinon continua as a distinguished feature of the spectral function in the liquid regimes, and are indicative of their robustness also in the thermodynamic limit.
The unambiguous detection of gapless phases, especially in the absence of definite Goldstone modes, remains challenging on the accessible systems. However, a thorough analysis of finite-size effects supports the conjectures about underlying theoretical descriptions of the phases considered: While the two-spinon bound state at $K$ in the Dirac spin-liquid regime constitutes a gapped minimum just like in the RPA approach~\cite{willsher_dynamics_2025}, the gap opens continuously with increasing $J_\chi$, leading via the hardly dispersive features in the chiral spin liquid to a distinct magnon branch in the tetrahedral order with the spectral intensity being reduced under this transition (Fig.~\ref{fig:spectral_cuts_KMX}\textbf{a}). The rotonlike minima at the $M$ points, by contrast, exhibit a finite-size gap in the DSL caused by the small cylinder circumference that is enhanced in the gapped chiral spin-liquid phase. 
Under the transition into the tetrahedral order, the gap at $M$ softens again to form (besides inevitable finite-size effects) Goldstone excitations. These phenomena are summarized in Fig.~\ref{fig:spectral_cuts_KMX} and---supplied with additional data for various points in the phase diagram---in Appendix~\ref{App:Overview}.
All our spectra from dynamical simulations find their confirmation and complement in the results obtained from the quasiparticle ansatz. In particular the overall distribution of spectral weight, despite possible finite-size effects in the numerical results, opens new doors to tell apart the various neighboring phases in experimental setups.
We could additionally identify subtle variations in spectral weight that are qualitatively supportive of the field-theoretical description of the chiral spin-liquid phase in terms of a Kalmeyer-Laughlin type wavefunction~\cite{Kalmeyer1989}. Even though these might be beyond experimental detectability in the nearby future, this aspect still demontrates notably the impact of powerful numerics that provide at least theoretically compelling insights into analytically mostly elusive phases. 

Our analysis compares comprehensively the effects of competing orders from adjacent phases on intermediate paramagnetic regimes, highlighting as a consequence the influence of quantum fluctuations on frustrated spins that exceed mean-field analysis or RPA-based approaches~\cite{willsher_dynamics_2025}.
A future route of research in this field could involve the investigation of higher-spin systems on discrete lattices. The technology that has become accessible invites to a discussion of multiple model systems on distinct lattices whose nature have been under long-standing debate.


\vspace{2mm}
\textit{Data availability}
The simulation data including plotting scripts are available via the zenodo repository at Ref.~\cite{zenodo2025}.
\vspace{2mm}

\section{Acknowledgements}
We thank Josef Willsher, Johannes Knolle, Sylvain Capponi and Andreas Läuchli for helpful discussions and their valuable comments and contributions.
FP acknowledges support by the Deutsche Forschungsgemeinschaft (DFG, German Research Foundation) under Germany’s Excellence Strategy EXC-2111-390814868, TRR360 (projectid492547816), and the Munich Quantum Valley, which is supported by the Bavarian state government with funds from the Hightech Agenda Bayern Plus. RM acknowledges support from  the Deutsche Forschungsgemeinschaft via the cluster of excellence ct.qmat (EXC 2147, project-id 390858490) and  SFB 1143 (Project-ID No. 247310070).

\bibliography{references_chiral_qsl}

\clearpage

\appendix

\onecolumngrid
\section*{Technical Details}
\twocolumngrid

\appendixtableofcontents

\appsection{Ground-State Optimization}
\label{App:GS}

We use the density matrix renormalization group algorithm for infinite systems (iDMRG) to obtain the ground state of a given model Hamiltonian by local optimizations. The wave function is expressed as a matrix product state (MPS) with a virtual maximum bond dimension $\chi$ and a physical Hilbert space dimension of $2$ for spin-$1/2$ systems. In order to represent the two-dimensional lattice, the system is mapped onto a cylinder with periodic boundary conditions along the translation vector ${\bf t} = L \cdot {\bf a}_2 - n\cdot {\bf a}_1$ corresponding to a geometry $\mathrm{YC}L-n$. For the ground-state optimization, we choose an infinite cylinder, i.e., the system has infinite boundary conditions along the direction of ${\bf a}_1$ realized by appropriate environments that are built up during the optimization sweeps~\cite{McCulloch2008}. In our case, we chose a unit cell with three rings of circumference $L$ for the iDMRG algorithm. Starting from some initial Néel product state, the total $S^z$ quantum number is conserved during the simulation. The data presented in the main text was generated for the geometry $\mathrm{YC}6$, hence the periodicity is closed by a translation along $L\cdot {\bf a}_2$ with $L=6$. In this Appendix, we discuss also dynamical data for other cylinder geometries such as $\mathrm{XC}6\equiv \mathrm{YC}6-3$, $\mathrm{YC}6-2$ and larger circumferences $L > 6$.

Whereas in the case of ordered phases, the iDMRG algorithm finds the desired ground state, the situation requires more care for the paramagnetic regimes where fractionalized excitations can lead to different topological sectors on cylinder geometries~\cite{CincioVidal2013, Zaletel2013, HeShengChen2014, Saadatmand2016}. For the spin-liquid phase in the $J_1-J_2$ Heisenberg antiferromagnet on the triangular lattice for $0.07\lesssim J_2/J_1 \lesssim 0.15$, for instance, there are two distinct sectors on cylinders $\mathrm{YC}L$ with even circumference $L$~\cite{Zhu2015, Hu2015, Hu2019}. In the picture of a resonating dimer-covering of the lattice, they are denoted as the even and odd sectors, depending on the parity of the number of dimers that are cut along the periodic circumference. The odd sector can be obtained either by adiabatically inserting a flux of $2\pi$ through the cylinder starting from the even sector that results directly from an iDMRG or DMRG simulation, or by enforcing the spinon excitations residing at the ends of the cylinder by cutting single sites at both boundaries from the system. Both methods yield equivalent ground states~\cite{Hu2015}. In the case of a finite system, the sites can be just removed at the open ends of the cylinder. For the iDMRG optimization we applied, we removed a single site at each boundary of the cylinder during the first few sweeps of iDMRG (or in general an odd number of sites for larger circumferences such as $L=12$). After these first sweeps, the full model system was restored for the remaining simulation. Due to the encoding of the boundary conditions in the environments for the infinite MPS, the algorithm yields the final ground state with the sought-after topological properties. The entanglement spectrum at bonds connecting rings of the cylinder exhibit a symmetric spectrum around the $s_z$ quantum number $s_z = -1/2$, indicating the spinon degrees of freedom at the boundaries~\cite{Hu2015, Saadatmand2016, Saadatmand2017}. In this work, we obtained all ground states for the $J_1-J_2$ spin-liquid regime using the method described, omitting single sites during the initial iDMRG sweeps to stabilize a ground state in the correct lowest-energy sector.

\begin{figure}
    \centering
    \includegraphics[scale=1]{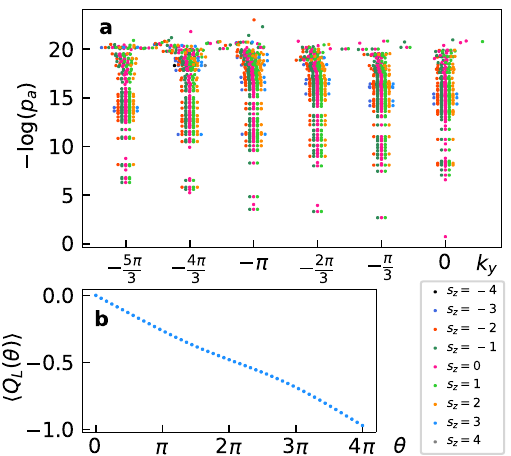}
    \caption{Momentum-resolved entanglement spectrum of the ground state at $J_\chi/J_1 = 0.2$ and $J_2/J_1 = 0.125$ at $\chi = 2000$ (\textbf{a}) and the average charge $\langle Q_L (\theta) \rangle$ at the left boundary of the cylinder under an adiabatic insertion of flux $\theta$ (the coupling parameters as before), which effectively represents the spin pumping through the cylinder (\textbf{b}). We observe that under the insertion of a flux of $2\pi$, one spinon with $s = 1/2$ is transported from one site of the cylinder to the other. The low-lying entanglement spectrum follows in the level counting the CFT prediction for each $s_z$ quantum number: 1, 1, 3, 5, 7, \ldots. For the sake of readability, the levels for finite $s_z$ have been shifted along the x-axis accordingly. The ground state has been obtained on an $\mathrm{YC}6$ infinite cylinder with a unit cell of three rings.}
    \label{fig:csl_entanglement_spectrum}
\end{figure}

The chiral spin liquid phase is characterized by gapless edge modes that manifest themselves in the momentum-resolved entanglement spectrum~\cite{LiHaldane2008, QiKatsuraLudwig2012, Gong2017, Szasz2020}. In Fig.~\ref{fig:csl_entanglement_spectrum}, we present the $k_y$-resolved entanglement spectrum in the chiral phase at $J_\chi = 0.2$ and bond dimension $\chi = 2000$. $p_a$ denote the eigenvalues of the reduced density matrix $\rho_A = \mathrm{Tr}_B\ket{\psi}\bra{\psi}$ for a bipartition of the system. In contrast to the momentum-resolved entanglement spectrum in the $J_1-J_2$ spin-liquid regime or ordered phases~\cite{Saadatmand2016, Saadatmand2017}, the entanglement spectrum is not symmetric around a certain $k_y$, but rather shows a chirality in the low-lying spectrum. The latter is separated from the generic higher part by a topological gap. The levels can be sorted by their $s_z$ quantum numbers. For small $s_z$ around $s_z = 0$, we observe a specific level counting that corresponds exactly to the prediction from a Wess-Zumino-Witten conformal field theory: 1, 1, 2, 3, 5, 7, \ldots~\cite{Wen1991, Szasz2020, MooreHaldane1997, CincioVidal2013}. Moreover, we can simulate a pumping experiment by adiabatically inserting a flux through the cylinder~\cite{Szasz2020}. The result is shown in Fig.~\ref{fig:csl_entanglement_spectrum}\textbf{\color{blue}b}: A flux of $\theta = 2\pi$ effectively pumps a spin-$1/2$ from the left boundary of the cylinder to the right-hand side. This serves as an illustration and confirmation of the fractionalized spinon excitations in the disordered regime.

Eventually, for the time evolution, we can use the iMPS ground states to build sufficiently large unit cells by periodically stacking them along the infinite ${\bf a}_1$ direction. This is decisive to avoid boundary effects from interferences and reflections when the perturbation signal reaches the boundaries of the system (see also the correlation spreading discussed in Appendix~\ref{App:Overview}). Since we are interested in the bulk physics, we can transform the large iMPS unit cell to a finite MPS for the sake of computational speed and stability. To do so, we obtain the dominant eigenvector of the transfer matrix to terminate the infinite system, and use segment DMRG to optimize over the outermost rings. This protocol has the advantage that various system sizes that are required for the dynamical simulations can be constructed from an iMPS ground state without the need to optimize the finite ground state for each system size starting from the initial state.

\appsection{Computing the Spectral Function}

\appsubsection{Numerical Protocol}
\label{App:Protocol}

\begin{figure}
    \centering
    \includegraphics[scale=1]{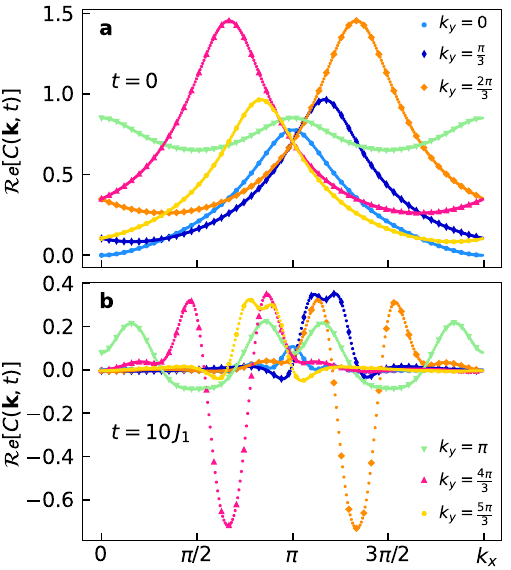}
    \caption{Illustration of the interpolation scheme in momentum space: We take the simulated values (big markers) and interpolate in $k_x$ on each cut in the Brillouin zone indicated by the different values of $k_y$. The interpolated values are plotted as smaller dots. We show the interpolation in the static structure factor (\textbf{a}) and after a finite evolution time (\textbf{b}). Real and imaginary part of $C({\bf k}, t)$ are interpolated separately. The data shown here is taken from the simulation in the Dirac spin liquid ($J_2/J_1 = 0.125$ with $\chi = 4000$).}
    \label{fig:momentum_interpolation}
\end{figure}

The core of our method consists of numerically computating the dynamical correlations $C({\bf k}, t)$ of the model considered in order to obtain the spectral function as the Fourier transform in spatial coordinates and time (cf. Eq.~\ref{equ:app_dsf}). However, due to the restrictions of our simulations regarding system size and evolution times---both being governed by the exponential increase in bond dimension $\chi$---, we need to apply certain numerical protocols to mitigate the emerging artifacts. First, in our numerical procedure we choose a time-step size of $\delta t = 0.04 \,J_1$ that allows us a sufficiently precise representation of the time-evolution operator $U(\delta t)$ following the MPO WII formulation that is suitable for long-range interactions as they occur in two-dimensional cylinder representations~\cite{Zaletel2015, Paeckel2019}. In fact, to reduce the error per time step from $\mathcal{O}(\delta t^2)$ by one order of magnitude, we apply two distinct operators in a row implementing complex time steps $\delta_1 = \frac{1+i}{2} \delta t$ and $\delta_2 = \frac{1-i}{2}\delta t$~\cite{Paeckel2019}. This is crucial to achieve a sufficient precision for the time evolution. The MPO WII algorithm often exhibits an artificial phase shift in energy per time step that needs to be corrected. Effectively, we evolve both the ground state and the perturbed state under identical conditions and finally obtain the dynamical correlations from computing the overlap 
\begin{equation}
    \bra{\psi_0(t)} \hat S^+_j \ket{\psi_{j_c}(t)}
\end{equation}
for the time-evolved ground state $\ket{\psi_0(t)}$ and perturbed state $\ket{\psi_{j_c}(t)}$ after the same number of time steps. In practice, we run two independent time evolutions and save the time-evolved MPS representations to the hard drive every $N_{\mathrm{STEPS}} = 5$ time steps. From these results, we can compute the correlations in time intervals $N_{\mathrm{STEPS}} \cdot \delta t$. To increase the time resolution, we interpolate the results to obtain the correlations in effective time steps of $\delta t$.

The transition from the spatial lattice coordinates to momentum space is performed via a discrete Fourier transformation that results in distinct cuts through the two-dimensional Brillouin zone according to the chosen geometry of the cylinder (see main text and Appendix~\ref{App:Overview} for illustrations). Since we can only simulation a finite time window, the direct Fourier transformation to the frequency regime would result in massive Gibbs oscillations. In order to reduce them, we first use a linear extrapolation scheme to extend the time series to a multiple of the actual simulation time $t_{\mathrm{sim}}$. (We use $m=20$ subsequent values to compute the entry at the following position.) Finally, we multiply the time series data with a Gaussian window function that is defined as acquiring the value $\alpha \ll 1$ at time $t_{\mathrm{sim}}$. Hence, all the extrapolated data will have only little weight in the final spectral function that arises from the convolution of the dynamical correlations in momentum space with the Gaussian signal with broadening $\sigma = \sqrt{t_{\mathrm{sim}}^2/(-2\ln(\alpha))}$.

Before the final Fourier transform to obtain the energy resolution, we apply a spline interpolation scheme for the momentum cuts individually (sorted by the transverse momenta $k_y$) for each time step. An demonstration of the applicability of this procedure is given in Fig.~\ref{fig:momentum_interpolation}. We present there the data obtained from the actual lattice of given dimensions (larger markers) and the interpolated data (small circles). This method thus allows us to increase the momentum space resolution for the spectral functions.

\appsubsection{Graphics Processing Units}
\label{App:GPU}

\begin{table}[h]
    \centering
    \begin{tabular}{c | c | c}
    \toprule
    Bond dimension &  Total time per step (sec.) & QR per step (sec.) \\ \midrule
    300 & 79 & 5.7 \\   
    1200 & 2570 & 111 \\  
    2000 & 6110 & 115 \\
    4000$^\ast$ & 27305 &  395 \\
    \bottomrule
    \end{tabular}
    \caption{Run-time estimates for the time evolution on CPUs. The bond dimension $\chi$ refers to the maximum virtual bond dimension of the MPS representation of the state. We use 16 CPU cores in parallel per simulation for all bond dimensions. The run times presented refer to a time evolution applying the MPO WII algorithm based on QR decompositions. The algorithm was applied to a ground state as obtained from DMRG for the candidate Dirac spin liquid at $J_2/J_1 = 0.125$ and $J_\chi = 0$ on a finite $\mathrm{YC}6$-cylinder with $L_x = 51$. The average total run time per time-evolution step has been rounded to seconds. For each simulation, we give an estimate of the average time needed for all QR decompositions that occur when performing a single time step. The actual run times might vary during the course of the simulation and are subject to uncertainties in particular when based on only few iterations ($^\ast$). We used Intel Xeon Platinum 8368 (3.4 GHz) and AMD EPYC 7763 (2.45 GHz) processors with Intel Distribution for Python version 3.9.16.}
    \label{tab:cpu_runtimes}
\end{table}

\begin{table}[h]
    \centering
    \begin{tabular}{c | c | c |c}
    \toprule
    Bond dim. & \# GPUs &  Time per step & QR per step \\ 
           &   & (sec.) & (sec.)\\ \midrule
    300 & 1 & 39  & 1.9 \\
    1200 & 1 & 55  & 2.3 \\
    2000$^\dagger$ & 1 & 248 & 4.7 \\ 
    2000 & 2 & 120 & 3.6 \\
    4000 & 4 & 668 & 335 \\
    \bottomrule
    \end{tabular}
    \caption{Run-time estimates for the QR time evolution on GPUs (NVIDIA A100 with 80 GB RAM each). We use the same algorithm as for Table~\ref{tab:cpu_runtimes}. We implement a memory management system that copies tensors between the GPU and system RAM, if necessary ($^\dagger$ cf. main text of the appendix for details on the implementation), depending on when they are needed on the GPU.
    We also implement a scheme to use multiple GPUs where each GPU memory holds part of the MPS and the corresponding environments: here used for $\chi = 2000$ (2 GPUs) and $\chi = 4000$ (4 GPUs). Sixteen CPU cores are used to perform the charge and block management of the tensors. This is done via compiled c-code. The number of cores used for this task hardly influences the overall runtime. All benchmark times for a full time-step update have been rounded to seconds.}
    \label{tab:gpu_runtimes}
\end{table}

\begin{figure}[h]
    \centering
    \includegraphics[scale=1]{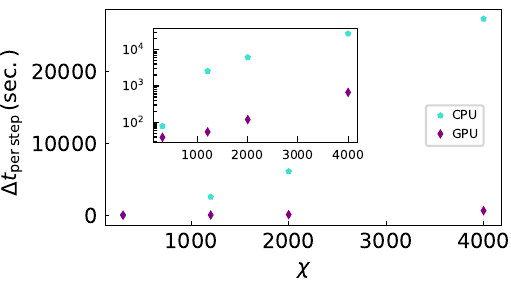}
    \caption{Illustration of the speedup in running the MPO WII time-evolution algorithm on CPU cores vs. GPUs. We plot the data presented in tables~\ref{tab:cpu_runtimes} and \ref{tab:gpu_runtimes}. The GPU run times refer to the optimal run times for the simulation in the candidate Dirac spin liquid using multiple GPUs if required by the memory consumption. The inset shows the same data with a log-scale for the run times in seconds.}
    \label{fig:speedup_gpu_vs_cpu}
\end{figure}

The variational application of the MPO from the WII approximation of the time-evolution operator can be achieved with various methods. We can use a two-site update that involves singular value decompositions (SVD). Alternatively, the variational application can be based on QR-decompositions that update single sites per step~\cite{Schollwoeck2011} or, in a slightly more involved scheme, can also achieve two-site updates~\cite{unfried_fast_2023}. In Appendix~\ref{App:Overview}, we compare the results from the two-site SVD update with the single-site QR implementation, thus demonstrating the equivalence of both algorithms for our model simulation. 
Tensor contractions and in particular QR decompositions have efficient and fast implementations on graphics processing units (GPUs)~\cite{li_numerical_2020, Pan2022, unfried_fast_2023}. This is advantageous for the run time of the time-evolution algorithm that we apply. Table~\ref{tab:cpu_runtimes} summarizes exemplary run times for the QR-based WII time evolution, applied to the ground state in the putative Dirac spin-liquid phase at $J_2/J_1 = 0.125$ and $J_\chi = 0$. For various bond dimensions, we state the ellapsed time in seconds for performing a single time-step. As outlined above, this comprises the application of two distinct matrix-product operators to the MPS. The actual run times achieved on a specific cluster may vary depending on the overall load of the machines and other exterior conditions. We also give an estimate of the computation time for the sum of all QR decompositions occurring for the $L\times L_x = 6\times 51$ system. We used 16 CPU cores in parallel. The processors used were of types Intel Xeon Platinum 8368 (3.4 GHz) and AMD EPYC 7763 (2.45 GHz).

Table~\ref{tab:gpu_runtimes}, in contrast, shows the results obtained for running the same algorithm for the identical model system on GPUs. We use a setup where a single node has 72 CPU cores with a RAM of 1 TB available. In addition to this, each node hosts four NVIDIA A100 GPUs with 80 GB RAM each. For the system size considered, we can run the full time evolution for suffciently small bond dimensions (up to $\chi = 1200$ in our example) on a single GPU NVIDIA A100. For $\chi = 2000$, however, the available GPU memory is not sufficient anymore. Note that only the data structures of the tensors from MPS and MPO are actually stored in the GPU memory. Since we implement tensors with Abelian $\mathbb{Z}_2$ or $U(1)$ symmetries, we have additional data to manage the charges of the tensor legs and define the positions of data blocks. This charge management is done completely on the CPUs and their RAM using compiled C-modules while the interface of the library is based on python.
To overcome the restrictions from the limited GPU memory, we implement various schemes: During the sweeps of the algorithm over the sites of the wavefunction, only a few tensors are actually needed in GPU memory at a certain point during the algorithm. Data that is currently not needed can be copied (as a background process) to the system RAM (and loaded again to the GPU when required). This allows us to run bond dimension $\chi = 2000$ also on a single GPU. The run time is still substantially faster than using the CPU backend. Note that even when using CPUs in parallel, there is a maximum in the run time for an intermediate number of CPUs since the parallelization routines need increasing resources for inter-core communication when the number of CPUs is augmented.
The alternative implementation---and usually, if resources are available, the preferred one---relies on combining multiple GPUs on the same node: A linearly subsequent part of the data structure (following the numbering of the sites) is loaded into each GPU-RAM. When performing the algorithmic sweep, tensors only have to be copied at the boundaries when the current center site jumps from one machine to the other. This protocol has been applied both to $\chi = 2000$ and $\chi = 4000$ with remarkable speedups compared to the other implementations.
Depending on the virtual bond dimension $\chi_{\mathrm{MPO}}$ of the MPO representing the Hamiltonian, and the system size, even the combination of up to four GPUs can turn out to not provide the required capacities. In this case, we used a combination of four GPUs and copied the tensors not currently needed (but as few as possible) between GPU-RAM and CPU-RAM forth and back. This approach has been used in the chiral phases.
Larger bond dimensions here would exceed the memory of the available CPU-RAM. In these cases, a new protocol would be possible that either recomputes data on the fly when required or uses the hard-drive disk to store data. This is a direction for future developments in the implementation of the tensor-network code.

Figure~\ref{fig:speedup_gpu_vs_cpu} illustrates the speedup gained from performing the time evolution on GPU infrastructure. The reduced run times result mainly from two effects: First, tensor contractions (and also QR decompositions) have very efficient implementations on GPUs. Second, operations on the GPU are performed in parallel if allowed by the dependencies of the data blocks among each other. 
This technology eventually opened new pathways for us to explore the dynamics in frustrated magnetic systems at high accuracy and for extended model Hamiltonians.

\clearpage

\onecolumngrid
\section*{Discussion of Dynamical Structure Factors}
\twocolumngrid

\appsection{Overview of the Spectral Data}
\label{App:Overview}
We present additional data for the spectral function of an extended Heisenberg model on the triangular lattice. The Hamiltonian is given as

\begin{align}
H &= J_1 \sum_{\langle i,j\rangle} \hat {\bf S}_i \cdot \hat {\bf S}_j
+ J_2 \sum_{\llangle i,j\rrangle} \hat {\bf S}_i \cdot \hat {\bf S}_j \nonumber \\
& + J_{\chi} \sum_{i,j,k \in \triangle, \triangledown} \hat {\bf S}_i \cdot \left(\hat {\bf S}_j \times \hat {\bf S}_k\right)\,.
\label{equ:app_Hamiltonian}
\end{align}

Using large-scale matrix-product-state simulations, we obtain the dynamical correlation function of the system and eventually the dynamical structure factor defined as

\begin{equation}
S^{+-}({\bf k}, \omega) = \hspace{-0.3em}\int \hspace{-0.3em} \mathrm{d}t  \sum_{j} e^{\ui\omega t - \ui{\bf k} \cdot ({\bf r}_j - {\bf r}_{j_c})} \braket{{\hat S}^{+}_j(t){\hat S}^{-}_{j_c}(0)}\,.
\label{equ:app_dsf}
\end{equation}

\begin{figure}
    \centering
    \includegraphics[scale=1]{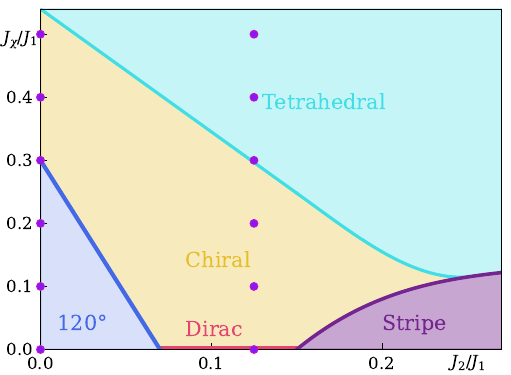}
    \caption{A schematic sketch of the phase diagram for the extended Heisenberg model on the triangular lattice following Refs.~\cite{Wietek2017, Gong2017}. The different colors indicate various ordered and liquid phases as denoted by the labels. The phase boundaries as depicted have been estimated using different numerical approaches~\cite{Wietek2017, Gong2017}. The phase boundaries drawn are hence understood to be approximations only. The purple dots mark the points in parameter space for which spectral data are presented in this Appendix.}
    \label{fig:phase_diagram_supplemental}
\end{figure}

In the following paragraphs, we give an overview of the spectral functions for various points in the parameter space of the model. These parameter sets are indicated as purple dots in the phase diagram Fig.~\ref{fig:phase_diagram_supplemental}. 

\appsubsection{$120^\circ$-Ordered Phase}
\label{app_subsec:120}

\begin{figure*}
    \centering
    \includegraphics[scale=1]{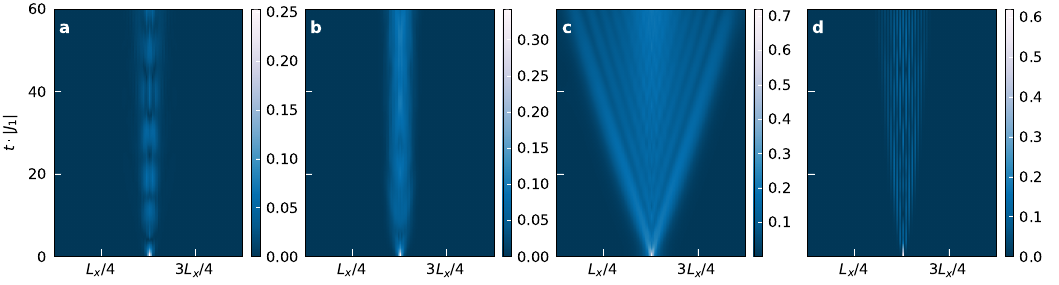}
    \caption{The real space correlations in the $120^\circ$-ordered phase at $J_2 = 0$ for $k_y$-resolved simulations with $k_y = 0$ (\textbf{a}), $k_y = \pi/3$ (\textbf{b}), $k_y = 2\pi/3$ (\textbf{c}), and $k_y = \pi$ (\textbf{d}). All simulations here were performed on a $YC6$-cylinder with $L_x = 126$ and bond dimension $\chi = 2000$. We clearly observe that the spreading velocity of the perturbation depends on the selected momentum cut. For the cut containing the Goldstone mode at $K$ (subplot \textbf{c}), the spreading is fastest, which corresponds to the large derivative of the dispersion when the magnon approaches the gapless point. Subplot \textbf{d} refers to the cut through $M^\prime-M$, which results in an intermediate spreading due to the cosine-like dispersion (cf. Fig.~\ref{fig:J20_convergence}).}
    \label{fig:J20_correlations_ky}
\end{figure*}

\begin{figure*}
    \centering
    \includegraphics[scale=1]{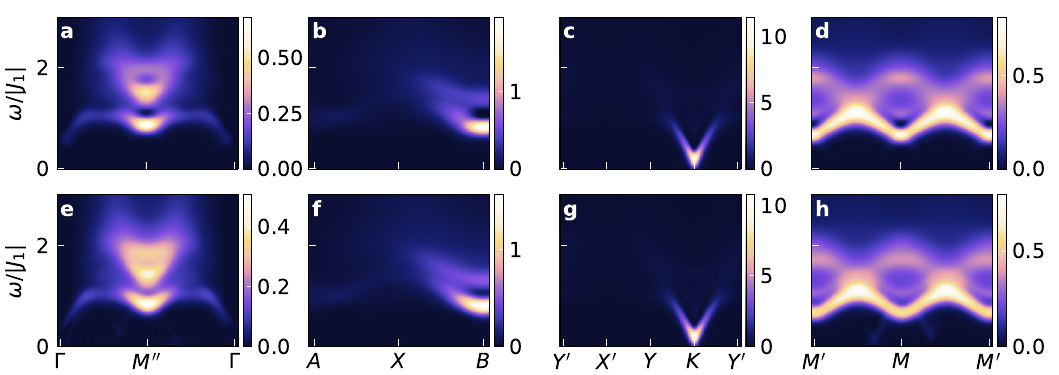}
    \caption{The spectral function at the Heisenberg point at $J_2=0$ from a time evolution with single-site perturbation in the system center (bottom row) for the parameters $\chi = 4000$, $L_x = 51$ compared with the results for $k_y$-resolved evolutions (top row) on a system size of $L_x = 126$ at maximum bond dimension $\chi = 2000$. The cylinder geometry used here is $YC6$, the evolution time for the data displayed $60 \,J_1$ ($\alpha = 10^{-11}$ defining the Gaussian window). We observe good agreement. For each subplot, the color scale is adjusted to the maximum of the spectral weight in this cut as obtained from the dynamical correlations.}
    \label{fig:J20_convergence}
\end{figure*}

\begin{figure}
    \centering
    \includegraphics[scale=1]{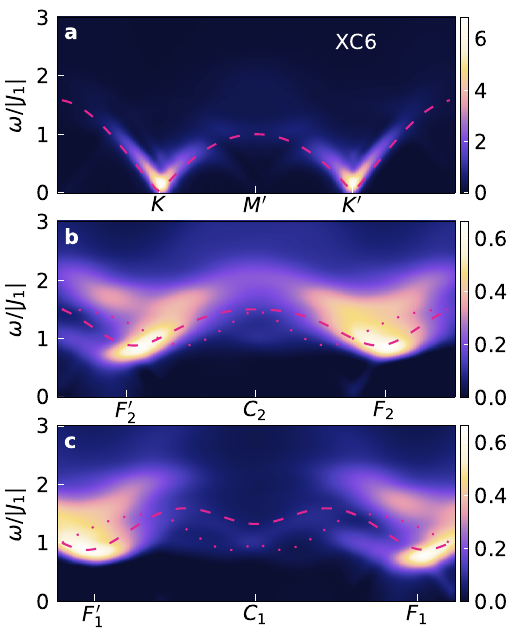}
    \caption{The dynamical spin structure factor at the Heisenberg point ($J_2 = 0$, $J_\chi = 0$) for a cylinder geometry $\mathrm{XC}6$ with system size $L \times L_x = 6 \times 51$. The simulations were performed with a virtual bond dimension $\chi = 2000$ for an evolution time of $60\,J_1$. The position of the momentum cuts and labels can be extracted from Fig.~\ref{fig:BZ_XC6_YC6_2}. In \textbf{a}, we clearly see the Goldstone modes with large spectral weight. The red dashed line denotes the dispersion from Linear Spin Wave Theory (LSWT). Panels~\textbf{b} and \textbf{c} show the cuts surrounding the $M$ point. The red dotted line indicates the lower bound of the two-magnon continuum as obtained from LSWT. We observe a mode with low spectral weight below the onset of the continuum. This illustrates the phenomenon of level repulsion as discussed in the main text.}
    \label{fig:J20_XC6}
\end{figure}

\begin{figure}
    \centering
    \includegraphics[scale=1]{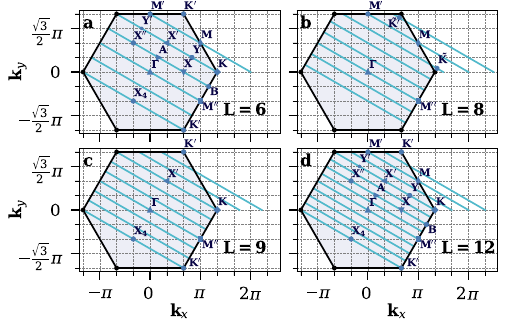}
    \caption{
    The accessible momentum cuts in the Brillouin zone for a cylinder of geometry $\mathrm{YC}L$ for various circumferences $L$. Relevant momentum points are explicitly named by a label.
    }
    \label{fig:BZ_YCL}
\end{figure}

At the Heisenberg point, i.e., $J_2 = 0$ and $J_\chi = 0$, the system is in the coplanar $120^\circ$ three-sublattice order. Figure~\ref{fig:J20_convergence} shows the dynamical spin structure factor for all available distinct momentum cuts on the $\mathrm{YC}6$-cylinder as illustrated in Fig.~\ref{fig:BZ_YCL}\textbf{\color{blue}a}. The bottom row Fig.~\ref{fig:J20_convergence}\textbf{\color{blue}e}-\textbf{\color{blue}h} summarizes the data from the main text that has been obtained on an $L\times L_x = 6\times 51$ cylinder at bond dimension $\chi = 4000$. We observe a sharp Goldstone mode at $K$ with a high concentration of spectral weight. Note that in this depiction, the colorbar in each subplot is adjusted to the corresponding maximum spectral weight along the momentum cut as represented in the subplot. For the figures in the main text, the color bar had been normalized to the total maximum of the spectral weight for a clearer comparison of the signals. In Fig.~\ref{fig:J20_convergence}\textbf{\color{blue}f}, the spectral intensity has not been cut off, which allows us to see only a faint hint of the low-intensity mode that is repelled from the two-magnon continuum. The renomalized dispersion with a rotonlike minimum at the $M$ points is visible in Fig.~\ref{fig:J20_convergence}\textbf{\color{blue}h}. Spectral weight at higher energies comprising a continuum and various modes occurs above the lowest single magnon branch. Fig.~\ref{fig:J20_convergence}\textbf{\color{blue}e} presents in addition to the discussion in the main text the spectral function on the cut involving the origin of the Brillouin zone, $\Gamma$. We detect a dispersive minimum at the momentum point $M^{\prime\prime}$ whose concentration of spectral weight ist comparable with the results at $M$ and $M^\prime$ (Fig.~\ref{fig:J20_convergence}\textbf{\color{blue}h}). A magnon branch softens as it approaches the $\Gamma$-point, becoming weaker in spectral intensity. At $\Gamma$ itself, the spectral weight vanishes completely as expected from the symmetry properties of the ground state. This is in accordance with the features of the static structure factor as discussed in the main text. The top row in Fig.~\ref{fig:J20_convergence} finally displays the same momentum cuts, but the spectral data has been obtained from momentum-resolved simulations, i.e., $k_y = 2\pi m/L$ ($m \in \{0, \dots L-1\}$ being an integer) has been fixed to a defined value during the time evolution. For details see the main text and its appendix. We observe in general good agreement between the single-site operator time-evolution data and the $k_y$-resolved case. Since the latter simulations have been performed on a cylinder with longitudinal dimension $L_x = 126$, the comparison demonstrates the robustness of the results under finite-size effects when varying $L_x$. Due to the different bond dimensions, however, there are visible differences in the convergence of the data: The momentum-resolved simulations have been performed with $\chi = 2000$. The dispersive mode at $B$, for instance, suffers in this case from truncation artifacts: The dispersion is less smooth than for $\chi = 4000$ (even though a $k_y$-resolved operator improves the accuracy of the simulation for otherwise identical simulation parameters). The horizontally pronounced strong signal at $B$ in Fig.~\ref{fig:J20_convergence}\textbf{\color{blue}b} is typical of an insufficient bond dimension. Similarly, the gap between the lowest branch and the onset of the continuum in Figs.~\ref{fig:J20_convergence}\textbf{\color{blue}d} and \textbf{\color{blue}h} is less pronounced in the case of the higher bond dimension (and correspondingly more accurate result).

Figure~\ref{fig:J20_correlations_ky} illustrates an interesting connection between the real-space simulation and the spectral dispersion: It shows the spatial time-dependent correlations at the Heisenberg point under the evolution of a momentum-resolved perturbation with different $k_y$. The steepest decent in the magnon dispersion is found at the gapless $K$ point. This results in a maximal spreading velocity of the perturbation as the derivative $\partial \varepsilon/\partial k$ of the mode yields the group velocity (Fig.~\ref{fig:J20_correlations_ky}\textbf{\color{blue}c}). The cosine-shaped dispersion on the cut $M$-$M^\prime$ results in an intermediate spreading (Fig.~\ref{fig:J20_correlations_ky}\textbf{\color{blue}d}), whereas the strongly pronounced, almost flat feature at $B$ produces barely any widening of the perturbed correlations at longer evolution times (Fig.~\ref{fig:J20_correlations_ky}\textbf{\color{blue}b}).

\begin{figure}
    \centering
    \includegraphics[scale=1]{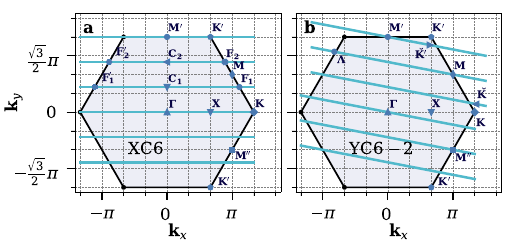}
    \caption{
    The accessible momentum cuts in the Brillouin zone for a cylinder of geometry, \textbf{a}, $\mathrm{XC}6$ and, \textbf{b}, $\mathrm{YC}6-2$.
    }
    \label{fig:BZ_XC6_YC6_2}
\end{figure}

Finally, we present the spectral function for the same phase, but on a different geometry: Figure~\ref{fig:J20_XC6} summarizes the data for an $\mathrm{XC}6$ cylinder with length $L_x = 51$. The bond dimension applied was $\chi = 2000$. Fig.~\ref{fig:J20_XC6}\textbf{\color{blue}a} refers to the momentum cut that includes a full side of the Brillouin zone comprising $K$, $M^\prime$, and $K^\prime$ (cf. Fig.~\ref{fig:BZ_XC6_YC6_2}\textbf{\color{blue}a}). We easily identify the gapless Goldstone modes that come along with a significant concentration of spectral weight. Departing from the $K$ points, the magnon branch would form a simple arch between the gapless points according to linear spin wave theory (LSWT) (cf. the red dashed line in Fig.~\ref{fig:J20_XC6}\textbf{\color{blue}a}). However, due to interaction effects and quantum fluctuations, the mode forms a faint rotonlike minimum at $M^\prime$. This is in accordance with the findings for the $\mathrm{YC}6$ cylinder as discussed in the main text. Figs.~\ref{fig:J20_XC6}\textbf{\color{blue}b} and \textbf{\color{blue}c} show the two cuts at equal distance to $M$ (Fig.~\ref{fig:BZ_XC6_YC6_2}\textbf{\color{blue}a}). The magnon branches have substantially less spectral weight than at the $K$ points. The lowest-energy signal follows closely the results from LSWT over most parts of the Brillouin zone as visualized by the red dashed line. However, in those regions where the onset of the two-magnon continuum lies lower in energy than the single-magnon branch---we indicate the lower bound of the continuum by the red loosely dotted line in Fig.~\ref{fig:J20_XC6}---there is spectral weight visible below the onset of the continuum. This is a signature of the level repulsion or avoided quasiparticle decay as also highlighted in the main text: Due to strong spin interactions, the lowest spin-wave mode does not simply enter the continuum where decay would be kinetically possible, but instead, it is at least partially repelled and forms a stable low-energy mode across the Brillouin zone~\cite{Verresen2019}.

\appsubsection{Candidate Dirac Spin Liquid}
\label{app_subsec:dsl}

\begin{figure*}
    \centering
    \includegraphics[scale=1]{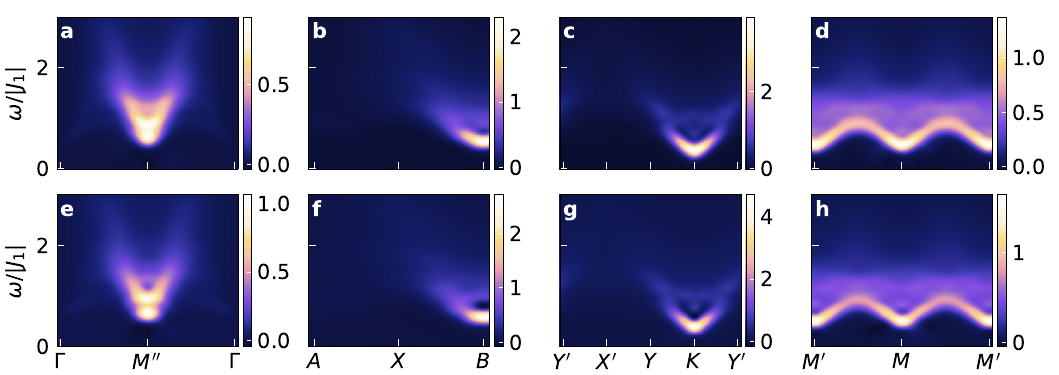}
    \caption{Convergence of the spectral function in the putative Dirac spin-liquid phase at $J_2/J_1=0.125$, $J_\chi=0$ for bond dimension $\chi=4000$ using the QR-based time-evolution algorithm (top row \textbf{a}-\textbf{d}) versus bond dimension $\chi=2000$ for the SVD-based formulation (bottom row \textbf{e} - \textbf{h}). All simulations were performed for a $YC6$-cylinder with extend $L_x = 51$. The maximum evolution time for the data plotted here is $60\,J_1$ at $\delta_t = 0.04\,J_1$. The Gaussian broadening was chosen with the parameter $\alpha = 10^{-8}$. We observe good agreement with only very few, largely suppressed artifacts even at long evolution times. The gap between the mode of bound spinons and the continuum above it vanishes for increased bond dimension (cf. \textbf{b} vs. \textbf{f} or \textbf{d} vs. \textbf{h} for instance). Longitudinal horizontal features often indicate truncation effects due to insufficient bond dimension (cf. \textbf{a} vs. \textbf{e}). The smoothness and resolution of the spinon-bound state improves for larger bond dimension (cf. \textbf{b} vs. \textbf{f}).}
    \label{fig:dsl_convergence}
\end{figure*}

At $J_2/J_1 = 1/8$ (with $J_\chi = 0$), we are at the classical phase transition point that is believed to be surrounded by an extended quantum spin liquid regime, potentially described by the algebraic Dirac spin liquid (DSL)~\cite{Hermele2005, Hu2019, Wietek2024}.
Figure~\ref{fig:dsl_convergence} shows the full spectral function for the $\mathrm{YC}6$ cylinder. We compare the high-resolution data for bond dimension $\chi = 4000$, obtained from a QR-based implementation of the time-evolution algorithm (top row), with the previously published data resulting from a SVD-based implementation~\cite{Drescher2023}, applied at $\chi = 2000$ (bottom row). Both simulations have been performed on a cylinder with length $L_x = 51$. We note that the general features of the spectrum are identical and seem well converged. However, there are subtleties that reveal the influence of the additional states with tiny Schmidt values that are kept during the evolution: The minima with an accumulation of spectral weight at $B$ and especially $K$ appear smoother and exhibit less of a flat horizontal feature that is characteristic of truncation effects. The same phenomenon can be seen at $M^{\prime\prime}$ between the $\Gamma$ points. Strikingly, the spinon continuum does not show any gap at all for the high bond dimension results, in particular along the cut $M^\prime$ - $M$ in 
Fig.~\ref{fig:dsl_convergence}{\color{blue}\textbf{d}}. The vanishing of spectral weight directly above the dispersive modes at $B$ and $K$ is also reduced in the more accurate simulation. The overall distribution of spectral weight stays roughly constant, with the intensification of weight at $K$ being slightly more balanced for $\chi = 4000$ than for $\chi = 2000$. This seems a general effect: Less accurate correlations for relatively long simulation times result in overemphasized maxima in the spectral intensity. For larger bond dimension, Gibb's oscillations are minimized.
These findings support the picture of a liquid phase with prominent spinon continua that come alongside strong quantum fluctuation beyond the parton-ansatz of DSL mean-field theory.

\begin{figure}
    \centering
    \includegraphics[scale=1]{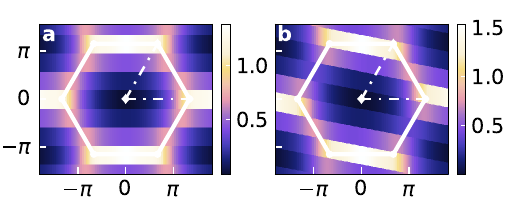}
    \caption{The static structure factor for $J_2/J_1=0.125$ ($J_\chi=0$) for the cylinder geometries, \textbf{a}, $\mathrm{XC}6$ and, \textbf{b}, $\mathrm{YC}6-2$. The MPS bond dimension is $\chi = 2000$.}
    \label{fig:ssf_XC6_YC6_2}
\end{figure}

\begin{figure}
    \centering
    \includegraphics[scale=1]{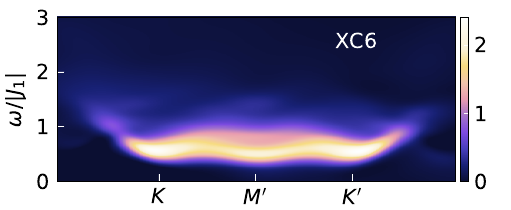}
    \caption{Dynamical structure factor in the candidate Dirac spin liquid phase at $J_2/J_1 = 0.125$ and $J_\chi=0$ for the geometry $\mathrm{XC6}$. The simulations have been performed on a system size of $L\times L_x = 6 \times 51$ with an MPS bond dimension $\chi = 2000$ for a maximum simulation time of $t_{\mathrm{sim}} = 60\,J_1$.}
    \label{fig:XC6}
\end{figure}

\begin{figure}
    \centering
    \includegraphics[scale=1]{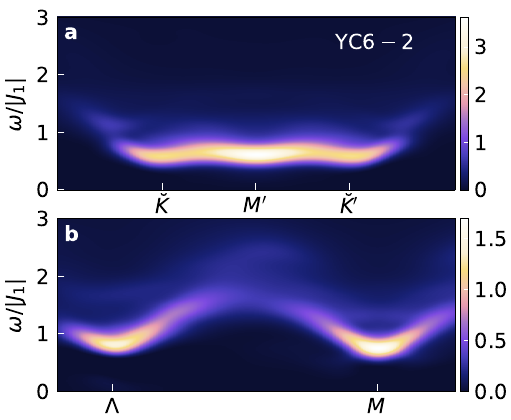}
    \caption{Dynamical structure factor in the candidate Dirac spin liquid phase at $J_2/J_1 = 0.125$ and $J_\chi=0$ for the geometry $\mathrm{YC6-2}$. The location of the momentum cuts can be read off from Fig.~\ref{fig:BZ_XC6_YC6_2}. The simulations have been performed on a system size of $L\times L_x = 6 \times 51$ with a maximum bond dimension $\chi = 2000$ for an evolution time of $t_{\mathrm{sim}} = 60\,J_1$.}
    \label{fig:YC6_2}
\end{figure}

In addition to the common $\mathrm{YC}L$ cylinder, we can also consider different geometries. Generally, the notation $\mathrm{YC}L-n$ means that the periodicity of the cylinder is closed by a translation of sites with the vector $L\cdot {\bf a}_2 - n\cdot {\bf a}_1$. Hence, $\mathrm{YC}L-L/2 \equiv \mathrm{XC}L$ denotes a cylinder layout where the cylinder circumference is closed exactly along the cartesian $\hat y$ direction for even $L$, following a zigzag ordering of sites as the most convenient and efficient choice. The resulting momentum cuts for two further geometries that we investigated, $\mathrm{XC}6$ and $\mathrm{YC}6-2$, can be read off from Fig.~\ref{fig:BZ_XC6_YC6_2}.
In Fig.~\ref{fig:ssf_XC6_YC6_2}, we show the resulting static structure factor for those two additional geometries. Both states have been optimized using iDMRG at $J_2/J_1 = 0.125$ ($J_\chi = 0$ throughout this paragraph) with a maximum bond dimension $\chi=2000$. Similarly to the $\mathrm{YC}6$ optimization, we applied an adiabatic flux insertion of $2\pi$ to reach the topological sector on the cylinder that has a more isotropic correlation structure. The energies per site are identical up to a precision of $10^{-7}$. We recognize a broad line of high spectral weight around the $M$ and $K$ points (that lie on one line for $\mathrm{XC}6$; for $\mathrm{YC}6-2$, the $K$ points are not hit exactly; we call the closest accessible momentum points $\breve{\bf K}$ and $\breve{\bf K}^\prime$). $M$ is only part of the set of accessible momenta in the latter case: it exhibits an intermediate maximum in spectral weight (similarly to the data for $\mathrm{YC}6$ as discussed in the main text). Due to the cylinder geometry and connected anisotropies, $M^{\prime\prime}$ is very low in spectral weight compared to the other $M$ points in this geometry. For $\mathrm{XC}6$, only $M^\prime$ is actually included (cf. Fig.~\ref{fig:BZ_XC6_YC6_2}). Up to the tild in the momentum cuts, the spectral weight distribution in the static structure factor looks comparable for both geometries, and the results are compatible with the findings from $\mathrm{YC}6$ cylinders that are subject of the main discussion of the article. The corresponding dynamical spin structure factors are presented in Figs.~\ref{fig:XC6}-\ref{fig:YC6_2}. The data from Fig.~\ref{fig:XC6} can be directly compared to Ref.~\cite{Sherman2023} where the same geometry at smaller bond dimension and system size has been investigated. In principal, the characteristics of the spectral functions agree between our results and Ref.~\cite{Sherman2023}. However, the longer evolution time of $60\,J_1$ renders finite-size effects such as the gaps at $M^\prime$ and $K$ unambiguously visible. Note that we do not apply a logarithmic scale for the spectral intensity. The spectral intensity at $M^\prime$ stays slightly under the one at $K$ and related points. The finite-size gaps at both momentum points are of comparable size. The same statement holds for the slightly tilded cut for $\mathrm{YC}6-2$ in Fig.~\ref{fig:YC6_2}, except for the spectral weight at $M^\prime$ being larger than at $\breve{\bf K}$. This is expected as the momentum cut has been rotated away from the exact $K$ point, and the signal having its maximum at $K$ indeed is thus consistent with our simulation outcomes. $M$ itself exhibits along this cut a nicely shaped rotonlike minimum, whose gap though appears even enhanced compared to $M^\prime$. We detect likewise a minimum in the two-spinon bound state at $\Lambda$ between $M^{\prime\prime}$ and $K$. Eventually, we can state that our findings for the various $L=6$ cylinders are consistent among each other. However, the small circumference $L$ seems to be the dominant factor to drive finite-size effects, which is why---up to differently emphasized cylinder anisotropies and artifacts of such kind--- the magnitude of the finite excitation gaps remain basically identical across the geometries considered even though the individual set of accessible momenta can vary significantly (possibly causing variations in the spectral weight for instance).

\begin{figure*}
    \centering
    \includegraphics[scale=1]{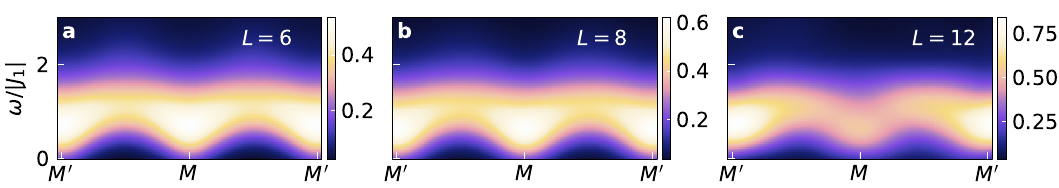}
    \caption{
    The spectral function in the Dirac spin liquid phase at $J_2/J_1=0.125$ along the cut $M^\prime$-$M$ for three different circumferences $L=6$ (\textbf{a}), $L=8$ (\textbf{b}), and $L=12$ (\textbf{c}). The total evolution time is for all subplots $t_{\mathrm{sim}} = 15\,J_1$ ($\alpha = 10^{-8}$). The simulations for $L=8$ and $L=12$ were performed at bond dimension $\chi=2000$ each, $L=6$ at $\chi=4000$. On all system sizes, we can reproduce the minima at the $M$-points that correspond to spinon-bilinear excitations in the theory of a Dirac spin liquid. However, in particular for $L=12$, the anisotropies in the spectral weight hint already at insufficient bond dimensions and dynamical correlations that are not converged yet. Whereas there is a trend for the spectral weight to increase with the circumference $L$, we need to assume that for $L>6$ the data may not have fully converged yet.
    }
    \label{fig:M_variousL}
\end{figure*}

\begin{figure*}
    \centering
    \includegraphics[scale=1]{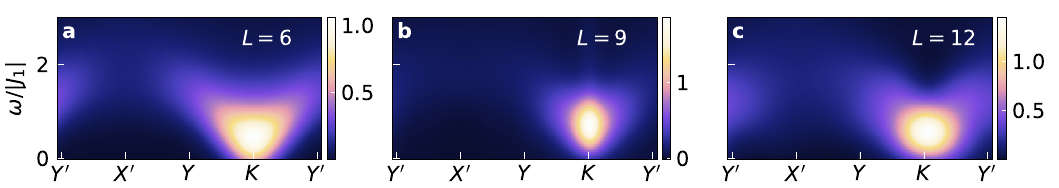}
    \caption{
    The spectral function in the Dirac spin liquid phase at $J_2/J_1=0.125$ along the cut $X^\prime$-$K$ for three different circumferences $L=6$ (\textbf{a}), $L=9$ (\textbf{b}), and $L=12$ (\textbf{c}). The simulation parameters are identical to those of Fig.~\ref{fig:M_variousL}. For all values of $L$, we can identify the minimum at strong spectral weight at $K$, which occurs at the points of the triplet monopole excitations. A similar spectral signal can be observed at $K^\prime$ (not shown in the panels presented in this work). For $L=9$ and $L=12$, the blob-like structure of the minima indicates that the dynamical correlations are affected by truncation effects due to the approximations of the time-evolution algorithm even despite the short evolution time.
    }
    \label{fig:K_variousL}
\end{figure*}

\begin{figure}
    \centering
    \includegraphics[scale=1]{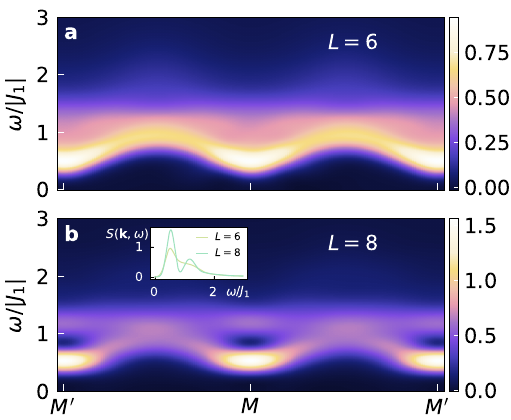}
    \caption{Comparison of the spectral function on the cut $M^\prime$~-~$M$ for the circumferences $L=6$ and $L=8$ at $J_2/J_1=0.125$ ($J_\chi=0$). The total simulation time here was $t_{\mathrm{sim}} = 40\,J_1$. We observe in both cases the characteristic minima at the $M$ points, whereas for $L=6$ the distribution of spectral weight is much more isotropic than for $L=8$. Moreover, the continuum above the spinon-bilinear excitation in the interpretation of the Dirac spin liquid is more pronounced and dense for $L=6$. Eventually, this hints at a better convergence for the smaller circumference. The inset in panel \textbf{b} shows the profiles of the spectral function at $M^\prime$ for both circumferences. The gap seems of equal size in both cases. The used virtual bond dimension for the matrix-product states were $\chi = 4000$ ($L=6$) and $\chi = 2000$ ($L=8$).}
    \label{fig:M_6_vs_8}
\end{figure}

\begin{figure}
    \centering
    \includegraphics[scale=1]{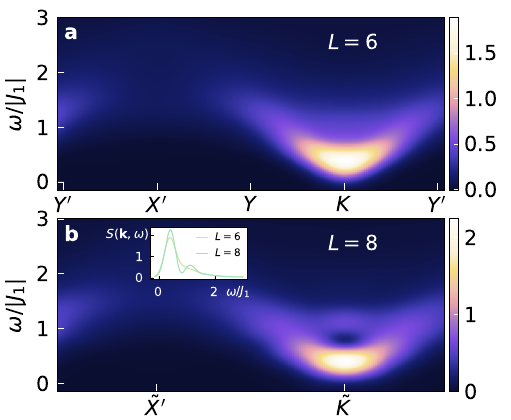}
    \caption{Comparison of the spectral function near the $K$~point for the circumferences $L=6$ and $L=8$ at $J_2/J_1~=~0.125$ ($J_\chi=0$). The total simulation time here was $t_{\mathrm{sim}}~=~30\,J_1$. Besides the discussed minimum at the $K$-point with the onset of a continuum above the mode, we observe a similar minimum at $\tilde K$ for $L=8$, which momentum point is closest to $K$ on the available cut. An identical result can be found for $K^\prime$ and $\tilde K^\prime$. Since for $L=8$, $K$ is not accessible, we expect not to see the absolute minimum of the mode in the corresponding spectral data. The inset in panel \textbf{b} show the profiles of the spectral function at $K$ for both values of $L$. 
    }
    \label{fig:K_6_vs_8}
\end{figure}

Finally, the enhancements from the GPU cluster infrastructure, especially with regard to the run times for intermediate bond dimensions of a few thousand Schmidt states, allowed us to run simulations for larger cylinder circumferences. To be precise, we performed simulations on $\mathrm{YC}L$ geometries for $L=8$, $L=9$ and $L=12$. As the complexity of the simulations increases exponentially with $L$, we could not reach the same evolution times as for the main data in this work. Nevertheless, the results presented here can give a glimpse of finite-size effects and scaling behavior, and illustrate the challenges and restrictions that have to be overcome in future computations.

Figure~\ref{fig:BZ_YCL} gives an overview of the Brillouin zone cuts as obtained for various circumferences. Specific momentum points that occur in the spectral functions presented are explicitly labeled. In Fig.~\ref{fig:M_variousL}, we show the cut $M^\prime$-$M$ for $L=6,8,12$. Due to the decay of the correlations during time evolution for insufficient bond dimension (we could only use $\chi = 2000$ because of RAM restrictions), we are confined to consider comparably short simulation times of $15\,J_1$. For the short evolution, the resulting broadening is huge and the minima in the spectrum at the $M$ points extend almost down to zero energy. Qualitatively, $L=6$ and $L=8$ yield equivalent results. $L=12$ produces a different distribution of spectral weight at $M^\prime$ compared to $M$. This is most likely due to truncation effects and not fully converged ground state correlations. We observe that the maximum of the spectral weight increases with $L$. This could be a result of insufficient bond dimension and convergence issues as well (see also the discussion above in Sec.~\ref{app_subsec:120}).
Figure~\ref{fig:K_variousL} displays the spectral function on the cut through the corner of the Brillouin zone for the same set of circumferences. The technical parameters are identical as for Fig.~\ref{fig:M_variousL}. The bound state at $K$ is not yet fully developed for this simulation time. In contrast, it appears as a broad, highly pronounced triangular feature that extends to vanishing energies at $K$ for $L=6$. For $L=9$ and $L=12$, the branches originating from the minimum are less refined yet. In these cases, the minimum is still existent, but it emerges as an elliptic accumulation of spectral weight. The maximum of the spectral intensity (which can serve as an indication of the finite-size gap) occurs at a higher energy value for $L=9$ than $L=6$. It softens slightly again for $L=12$. Though there are presumably still strong effects from convergence issues, the discussed features seem to suggest that even up to $L=12$, the small circumference imposes a variety of finite system-size restrictions that prevent us potentially from extracting clear scaling behaviors even for highly resolved simulations. Similar findings have been reported for the analysis of the transfer-matrix spectrum~\cite{Hu2019}.

Concentrating on reliably converged dynamical data, we compare $L=6$ and $L=8$ for two different regions in the Brillouin zone in Figs.~\ref{fig:M_6_vs_8} and \ref{fig:K_6_vs_8}. We could go now to longer simulation times of $30\,J_1$ and $40\,J_1$ respectively. Along the cut $M^\prime$-$M$, we find good agreement in the overall features. The onset of the continuum for $L=8$ above the $M$ points displays a gap between lowest mode and continuum which disappears for the spectral function for $L=6$ at high bond dimension. We would expect this to be an artifact caused by the requirement to use a much larger bond dimension for $L=8$ in order to achieve the same level of precision. The profiles in the inset show that the location of the maximum spectral intensity is situated at the same energy. As discussed before, the finite-size effects don't seem to be affected much by changing the cylinder circumference within the restricted set of the accessible small integers. Interestingly, the minimum at $M$ collects for $L=8$ again a profoundly higher spectral weight that in the well-converged $L=6$ case. 
Eventually, we compare the spectral function around the $K$ point for $L=6$ and $L=8$ in Fig.~\ref{fig:K_6_vs_8}. Note that the $K$ momentum is not accessible for $L=8$ that is not divisible by three. Instead, we denote the momentum point at minimal distance to $K$ as $\tilde K$. The minimum is well pronounced at $\tilde K$ as well, with the main difference to $K$ at $L=6$ being the vanishing spectral weight that emerges directly above the maximum. This is arguably a finite-accuracy feature. The inset depicts again the spectral profiles at $K$ and $\tilde K$ for the respective circumferences. The corresponding energy for the maximum value agrees among both cases. Since $\tilde K$ is slightly shifted compared to the expected minimal mode at $K$, this indicates that the gap at $K$, if accessible, would actually soften on this system size compared to $L=6$.

\appsubsection{Phases with Chiral Interaction}

\appsubsubsection{Chiral Spin Liquid}

\begin{figure}[H]
    \centering
    \includegraphics[scale=1]{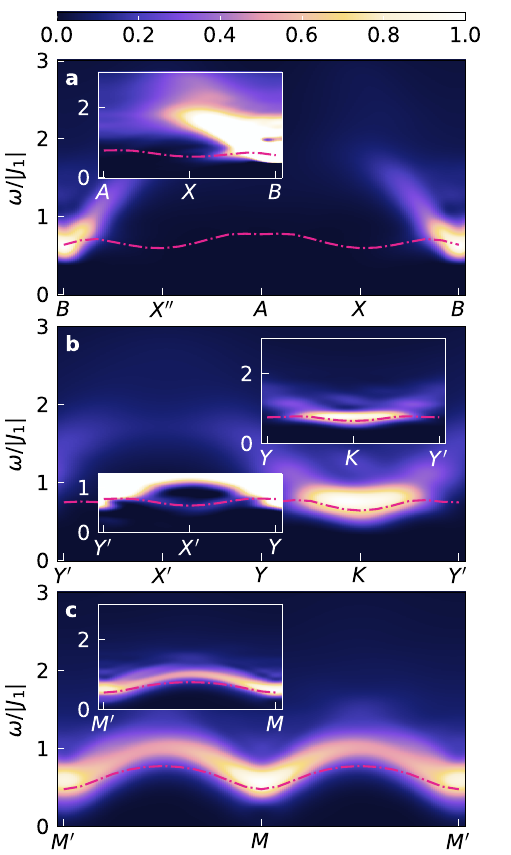}
    \caption{Spectral function in the chiral spin liquid at $J_\chi/J_1 = 0.1$ and $J_2/J_1 = 0.125$. In contrast to Fig.~{\color{blue}5} in the main text, we display exclusively data obtained from $k_y$-resolved simulations for $\mathrm{YC}6$-cylinders with length $L_x = 51$. The bond dimension is $\chi = 2000$. The main panels are as before for a Gaussian convolution with $\alpha = 10^{-11}$ and simulation time $t_{\mathrm{sim}} = 60\,J_1$, the insets with $\alpha = 10^{-6}$. The inset in \textbf{a} and the lower inset in \textbf{b} apply a cutoff of the spectral at ten percent of the maximum. The spectral functions from these simulations agree with the results obtained via a single-site perturbation (Fig.~{\color{blue}5} in the main text). The deviation of the curve with the largest spectral weight along the cut $M^\prime$-$M$ from the quasiparticle ansatz (red dot-dashed line) that we discussed in the main article is confirmed. The shape of the continuum along this cut is not fully symmetric around the center between $M^\prime$ and $M$, which most likely can be attributed to a finite-cylinder effect. The lower inset in \textbf{b}, by contrast, shows conclusive agreement between the lowest-energy mode from the quasiparticle optimization and the faint low-energy signal in the full spectral function. The inset comprising the $K$ point features a sharper Gaussian broadening and thus the quasiparticle mode clings tightly to the lower edge of the mode in the spectral function.}
    \label{fig:JX01_supplemental}
\end{figure}

\begin{figure}[H]
    \centering
    \includegraphics[scale=1]{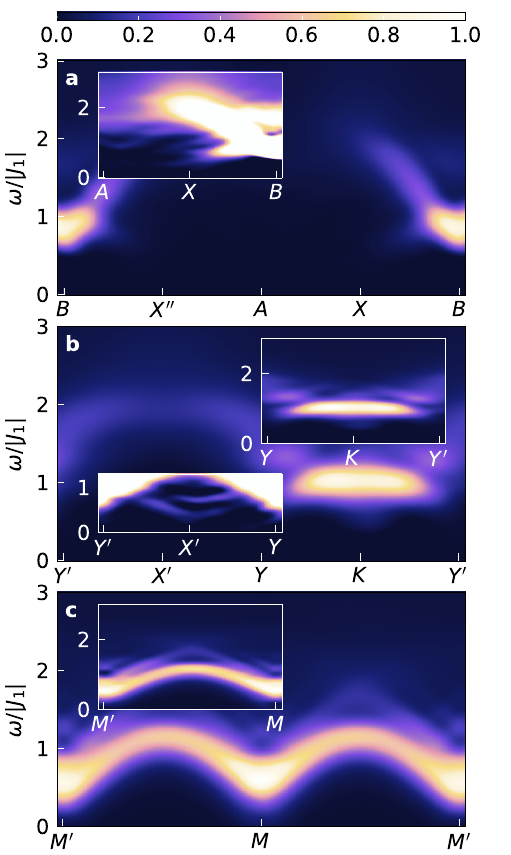}
    \caption{Spectral function deep in the chiral spin-liquid phase at $J_\chi /J_1 = 0.2$ and $J_2/J_1 = 0.125$. The data has been obtained for a geometry $\mathrm{YC}6$ with $L_x = 81$. The momentum cuts containing the $K$ point and the inset in \textbf{c} show momentum-resolved spectral functions ($L_x = 51$). The maximum bond dimension for all simulations has been set to $\chi = 2000$. The parameters for the spectral function are as in Fig.~\ref{fig:JX01_supplemental}. The broad signal and continuum along the cut $M^\prime$-$M$ is comparable to the case $J_\chi/J_1=0.1$ (Fig.~\ref{fig:JX01_supplemental}). The lower inset in \textbf{b} and the inset in \textbf{a} shows distinctly the faint low-energy spectral weight that forms a minimum at the $X$ points. Moreover, there is a weak distribution of spectral weight at the $K$ point below the dominant signal that forms as well an energetic minimum at $K$. This is in line with the discussion in the main text on the connection to the spinon-continuum from the Kalmeyer-Laughlin ansatz~\cite{Kalmeyer1987}.}
    \label{fig:JX02_supplemental}
\end{figure}

Figures~\ref{fig:JX01_supplemental} and \ref{fig:JX02_supplemental} show the spectral function in the center region of the chiral spin liquid at $J_\chi/J_1 = 0.1$ (cf. Fig.~{\color{blue}5} in the main text) and $J_\chi/J_1 = 0.2$ ($J_2/J_1 = 0.125$ in both cases). We observe a broad signal and continua in the cut along $M^\prime$-$M$ with minima in energy at the $M$ points (that, however, come along with maxima in the spectral intensity). The lower insets in Fig.~\ref{fig:JX01_supplemental}\textbf{\color{blue}b} and Fig.~\ref{fig:JX02_supplemental}\textbf{\color{blue}b} illustrate the occurrence of faint low-energy spectral weight below the dominant contributions that acquires the minimal energy at the momentum point $X^\prime$. This is in accordance with the findings discussed in the main text that bring these weak low-energy features in con-\linebreak nection to a minimum in the two-spinon continuum as it arises from the Kalmeyer-Laughlin wavefunction for the chiral spin liquid~\cite{Kalmeyer1987}.
A similar phenomenon is visible at $X$ in the insets in the upper panels. By increasing the chiral coupling constant from $J_\chi/J_1 = 0.1$ to $J_\chi/J_1 = 0.2$, the main signal around $K$ becomes more flat while in the latter case, a faint distribution of spectral weight forms a cone-like shape below the high-intensity mode with a minimum at $K$ itself. This supports the previous conjecture of a spinon-continuum that can be described via the Kalmeyer-Laughlin wavefunction ansatz. The single-spinon dispersion exhibits minima at the $X$ points in their evaluation~\cite{Kalmeyer1989}. The resulting minima in the two-spinon continuum would reside at $X$ points, $K$ points and the $M$ points. Also at the latter momenta, we observe a broad signal with spectral weight extended to low energies. A more detailed investigation of the spectral profiles is given in the main text with regard to Fig.~{\color{blue}7} and in Section~\ref{sec:convergence_supp} in this Appendix.

\begin{figure*}
    \centering
    \includegraphics[scale=1]{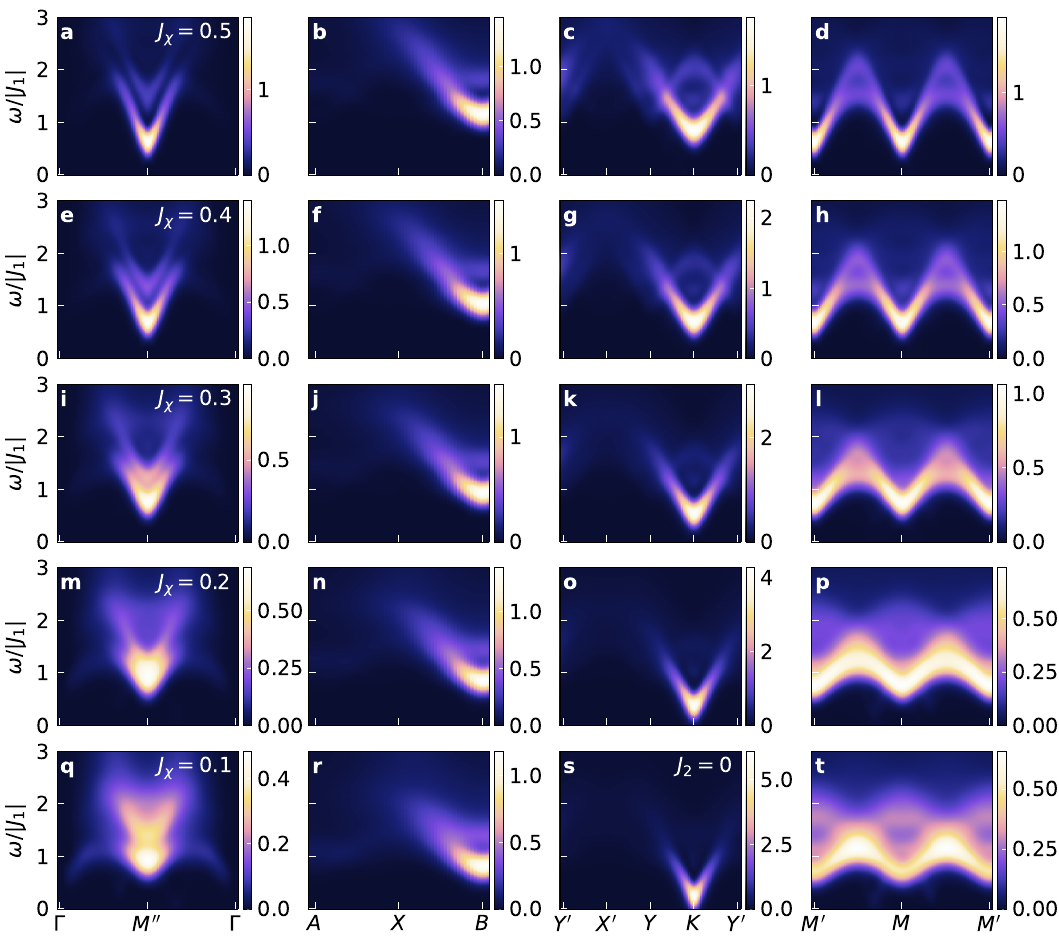}
    \caption{Spectral function for a sweep in $J_\chi \in [0.1, 0.5]$ at the Heisenberg point $J_2=0$. The simulations were performed on a $\mathrm{YC}6$ cylinder with $L_x = 51$ and MPS bond dimension $\chi = 2000$. As in the main plots the Gaussian window was defined via $\alpha = 10^{-11}$. The evolution time is $t_{\mathrm{sim}} = 40\,J_1$.}
    \label{fig:sweep_J20}
\end{figure*}

\begin{figure*}
    \centering
    \includegraphics[scale=1]{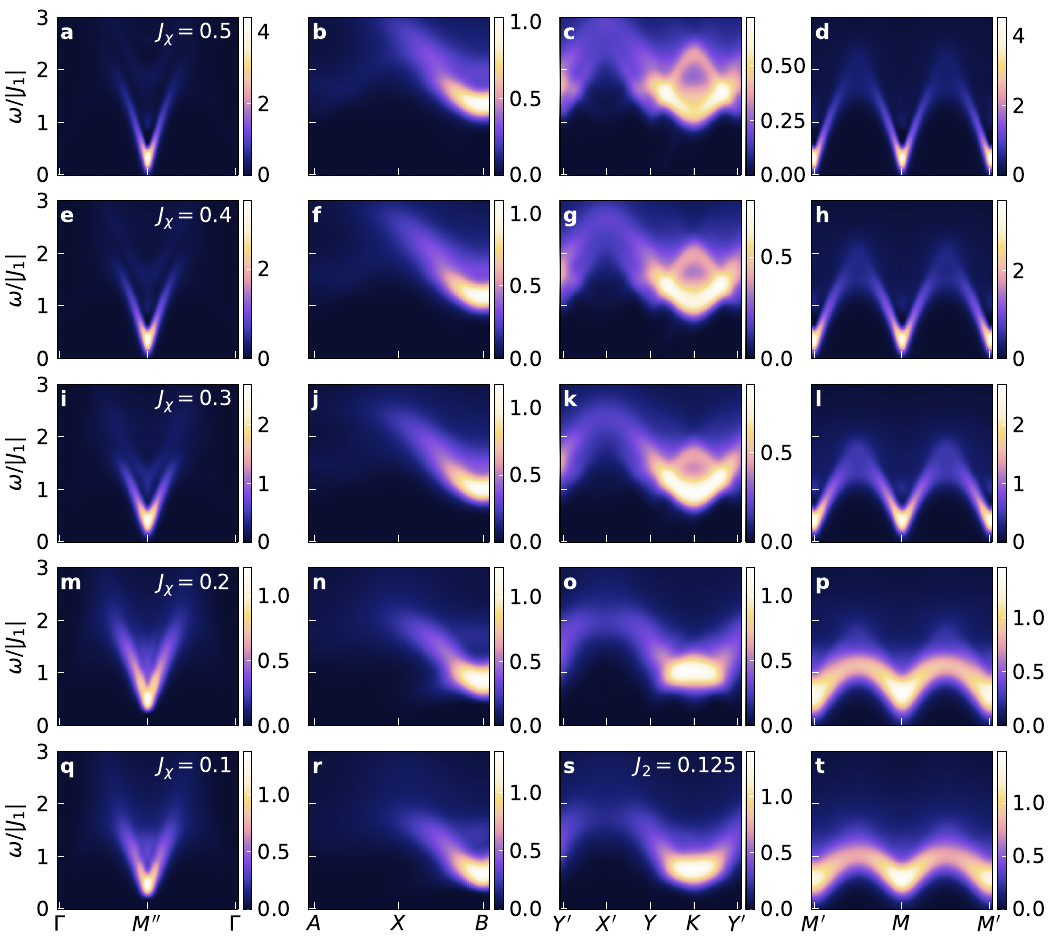}
    \caption{Spectral function for a sweep in $J_\chi\in [0.1, 0.5]$ at $J_2/J_1 = 0.125$. All data has been obtained for the geometry $\mathrm{YC6}$ with $L_x = 81$ and bond dimension $\chi = 2000$. As in Fig.~\ref{fig:sweep_J20}, $\alpha = 10^{-11}$ and $t_{\mathrm{sim}} = 40\,J_1$.}
    \label{fig:sweep_dsl}
\end{figure*}

\appsubsubsection{Sweeps in $J_\chi$}
To conclude this extensive discussion of spectral data for the extended Heisenberg model on the triangular lattice, we present in Figs.~\ref{fig:sweep_J20} and \ref{fig:sweep_dsl} the dynamical structure factor on all reachable momentum cuts for the $\mathrm{YC}6$-cylinder for a sweep in $J_\chi$ at $J_2 = 0$ and $J_2/J_1 = 0.125$ respectively.
The dispersive minimum at $B$ seems to be actually a stable feature of the spectral function throughout the considered parts of the phase diagram. This can be seen in the overview panels for the two series of chiral interactions $J_\chi = 0.1, 0.2, \dots, 0.5$ starting from the $120^\circ$ phase (Fig.~\ref{fig:sweep_J20}) and the Dirac spin liquid (Fig.~\ref{fig:sweep_dsl}).
At $M^{\prime\prime}$ in the ordered phase, we find the rotonlike minimum with an only intermediate spectral intensity. Fainting single magnon branches extend towards the $\Gamma$ points and soften on their path toward the center of the Brillouin zone (Figs.~\ref{fig:sweep_J20}\textbf{\color{blue}m} and \textbf{\color{blue}q}). In the chiral spin-liquid phase, coming from the DSL, the same momentum cut results in a sharp minimum at $M^{\prime\prime}$ with\hfill a\hfill smaller\hfill gap\hfill than\hfill in

\begin{figure}
    \centering
    \includegraphics[scale=1]{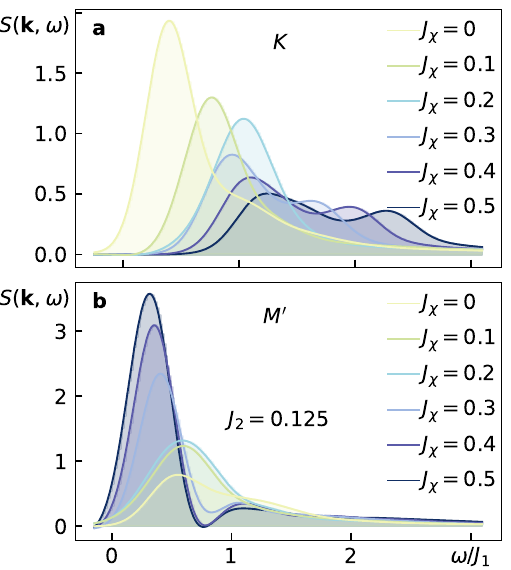}
    \caption{Spectral profiles of the momentum points $K$ and $M^\prime$ under the increase of $J_\chi$ starting from the Dirac spin liquid at $J_2/J_1 = 0.125$. The data plotted is compatible with Fig.~{\color{blue}7} in the main text ($\mathrm{YC}6$ cylinders): For the DSL, we used $\chi = 4000$ at $L_x = 51$, the data with finite chiral interaction $J_\chi$ has been obtained for $\chi = 2000$ and $L_x = 81$. The profile at $M^\prime$ and $J_\chi/J_1 = 0.5$ results from a $k_y$-resolved simulation with $\chi = 2000$ and $L_x = 51$. The evolution time underlying here is $t_{\mathrm{sim}} = 36\,J_1$, the Gaussian window chosen as $\alpha = 10^{-11}$ (the Goldstone mode at $M^\prime$ in the tetrahedral phase being delicate to resolve). We observe an opening of the gap at both momentum points in the gapped chiral spin-liquid phase. At $K$, the gap opens further after the transition to the tetrahedral phase ($J_\chi/J_1 \gtrsim 0.3$), whereas at $M^\prime$ it softens since the mode at this momentum forms a Goldstone mode under the tetrahedral order.}
    \label{fig:peak_shift_DSL_K_M}
\end{figure}

\begin{figure}
    \centering
    \includegraphics[scale=1]{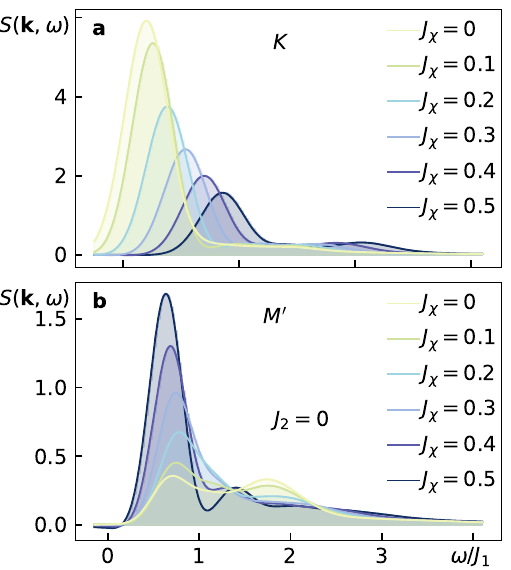}
    \caption{Spectral profiles at $K$ and $M^\prime$ evolving under increasing $J_\chi$ when starting from the $120^\circ$ order at $J_2 = 0$. All data was obtained on $\mathrm{YC}6$-cylinders. In the case of the $120^\circ$ phase at $J_\chi = 0$, we plot simulation data with $\chi = 4000$ and $L_x = 51$. The data for finite $J_\chi$ has been obtained for $\chi = 2000$ and $L_x  = 51$. As in Fig.~\ref{fig:peak_shift_DSL_K_M}, we observe an opening of the gap at $M^\prime$ while approaching and passing through the chiral spin liquid with a softening towards the transition to the tetrahedral phase ($J_\chi/J_1 \approx 0.5$). The profiles at $K$ show a continuous increase of the gap from the $120^\circ$ order across the chiral spin-liquid phase to the noncoplanar four-sublattice order.}
    \label{fig:peak_shift_120_K_M}
\end{figure}

\noindent the ordered phase for $J_2 = 0$. Coming from the supposedly gapless point in the Dirac phase, there is a slight opening of the gap in the chiral spin liquid (cf. Figs.~\ref{fig:sweep_dsl}\textbf{\color{blue}m} and \textbf{\color{blue}q} and in particular Fig.~\ref{fig:peak_shift_DSL_K_M} as well as the discussion in the main text). We observe for both values of $J_2$ first an opening of the gap at the $M$ points under turning on the chiral interaction. Finally, a sharpening and softening of the mode at $M^{\prime\prime}$ occurs when increasing $J_\chi$ further, moving towards the tetrahedral order that comes along with Goldstone modes at the $M$ points. This is illustrated in Figs.~\ref{fig:sweep_J20}-\ref{fig:peak_shift_120_K_M}.  In the case of $J_2/J_1 = 0.125$, $J_\chi = 0.4$ and $J_\chi = 0.5$ are expected to lie already within the tetrahedrally ordered phase. This is supported by the (except for finite-size effects) gapless magnon modes that we can detect at all $M$ points (Figs.~\ref{fig:sweep_dsl}\textbf{\color{blue}a}, \textbf{\color{blue}d}, \textbf{\color{blue}e}, and \textbf{\color{blue}h}). $J_\chi = 0.3$ is around the transition point towards the ordered phase: We observe already softened minima at the $M$ points with a significant reduction of continua as they arise from the spinon excitations in the chiral spin liquid in Figs.~\ref{fig:sweep_J20}\textbf{\color{blue}m}, \textbf{\color{blue}p}, \textbf{\color{blue}q} and \textbf{\color{blue}t}. Simultaneously, the weight at the minima at $M$, $M^\prime$, $M^{\prime\prime}$ is increased with the modes being less broad (compare Fig.~\ref{fig:sweep_dsl}\textbf{\color{blue}l} and \textbf{\color{blue}p} for an illustration of this aspect). 
On the contrary, the parameter points with $J_2 = 0$ have not reached the tetrahedral order yet for $J_\chi/J_1 = 0.3$, $J_\chi/J_1 = 0.4$ or even $J_\chi /J_1 = 0.5$. Whereas the middle value of those is expected to be located within the extended liquid regime, the other two values should mark approximately the transition points towards the adjacent ordered phases, namely $120^\circ$ and tetrahedral order. Increasing $J_\chi$, we observe an increase of spectral weight at the minimum at the $M$ points. In addition to this, the modes become more sharply defined. For $J_\chi/J_1 = 0.5$, we detect a slight softening of the gap at $M$ (Fig.~\ref{fig:sweep_J20}\textbf{\color{blue}d} vs. \textbf{\color{blue}h}). This is in accordance with the picture of a gapless Goldstone mode in the tetrahedral phase beyond $J_\chi/J_1 = 0.5$.
The shape of the continua in the chiral spin liquid phase are worth a comment as well: Whereas the spinon continuum has a rather flat higher energy structure for the DSL (Fig.~\ref{fig:dsl_convergence}\textbf{\color{blue}d}) and the magnon continuum seems to stretch out to the highest energy values above $M$ directly (Fig.~\ref{fig:J20_convergence}\textbf{\color{blue}h}, Fig.~\ref{fig:sweep_J20}\textbf{\color{blue}p} and~\textbf{\color{blue}t}), in the chiral spin liquid, the spinon continua appear more pronounced in the middle of the line $M$-$M^\prime$: This can be seen in Figs.~\ref{fig:sweep_dsl}\textbf{\color{blue}p} and~\textbf{\color{blue}t} or also in Figs.~\ref{fig:sweep_J20}\textbf{\color{blue}d}, \textbf{\color{blue}h} and \textbf{\color{blue}l}. Interestingly, for $J_2 = 0$ and starting from $J_\chi/J_1 = 0.3$ and larger, the previously broad spectral feature at $M^{\prime\prime}$ detaches into a pronounced low-energy mode and a more faint second mode that resides at higher energies and exhibits a sharp minimum as well. With increasing $J_\chi$, the lower bound state softens while the upper mode enhances the gap between the two by moving to higher energies.

Some of the key signatures of the proximate phases, however, can be identified at the $K$ point. Figure~\ref{fig:sweep_J20} shows that for $J_\chi/J_1 = 0.1$ and $J_\chi/J_1 = 0.2$, there is a strongly-pronounced low-energy mode with minimum at $K$ that is linearly approached from both directions. This feature we can identify as the gapless Goldstone mode that characterizes the coplanar $120^\circ$ order (Figs.~\ref{fig:sweep_J20}\textbf{\color{blue}s} and \textbf{\color{blue}o}). Starting from $J_\chi/J_1 = 0.3$ (Fig.~\ref{fig:sweep_J20}\textbf{\color{blue}k}), the gap at $K$ increases significantly. In the gapped chiral spin liquid phase, we observe a gapped minimum with reduced spectral weight and more developed continua (Fig.~\ref{fig:sweep_J20}\textbf{\color{blue}g}). Around the transition to the tetrahedral order ($J_\chi/J_1 = 0.5$ in Fig.~\ref{fig:sweep_J20}\textbf{\color{blue}c}), the gap opens further and signatures of single magnon branches emerge from the continuum at energies above the lowest mode. The Goldstone modes in the tetrahedral phase reside at the $M$ points. The single magnon branch at $K$ is gapped. Consequently, the resulting gap is more developed in Fig.~\ref{fig:sweep_J20}\textbf{\color{blue}c} compared to Figs.~\ref{fig:sweep_J20}\textbf{\color{blue}a} and \textbf{\color{blue}d}.  
In accordance with these findings, the spectral function for $J_\chi/J_1  = 0.3-0.5$ at $J_2/J_1 = 0.125$ exposes similar magnon branches as at $J_\chi/J_1 = 0.5$ with $J_2 = 0$. The spectral weight of these modes, however, is lower. From Fig.~\ref{fig:sweep_dsl}\textbf{\color{blue}k} over \textbf{\color{blue}g} to \textbf{\color{blue}c}, we observe an overall upward shift of the magnon dispersion. Higher-energy modes are emerging more distinctly when regarded deeper in the tetrahedrally ordered phase. These magnon structures stand in clear contrast to the flat and more pronounced spectral feature at $K$ in the chiral spin liquid phase for $J_2/J_1 = 0.125$ and $J_\chi/J_1 = 0.1$ and $J_\chi/J_1 = 0.2$. 
These very horizontal key signatures evolve in the center of the chiral spin liquid phase (Fig.~\ref{fig:sweep_dsl}\textbf{\color{blue}o} and \textbf{\color{blue}s}). Towards the boundaries, the minima are more dispersive, reflecting the proximity to the Goldstone modes of the $120^\circ$ order (Fig.~\ref{fig:sweep_J20}\textbf{\color{blue}g}).

Figures~\ref{fig:peak_shift_DSL_K_M}-\ref{fig:peak_shift_120_K_M} illustrate the main shifts of spectral intensity at $K$ and $M^\prime$ when increasing $J_\chi$ starting from both the $120^\circ$ order and the DSL, and the evolution of the dectected gaps at both momentum points. Both features, as outlined above, can be related to the underlying ordered and liquid phases.


\appsection{Convergence of Spectral Functions}
\label{sec:convergence_supp}

In this section, we present a more detailed discussion on the convergence of the spectral function at selected momentum points that play a crucial role in the interpretation of the data with respect to the underlying theories.

\begin{figure}
    \centering
    \includegraphics[scale=1]{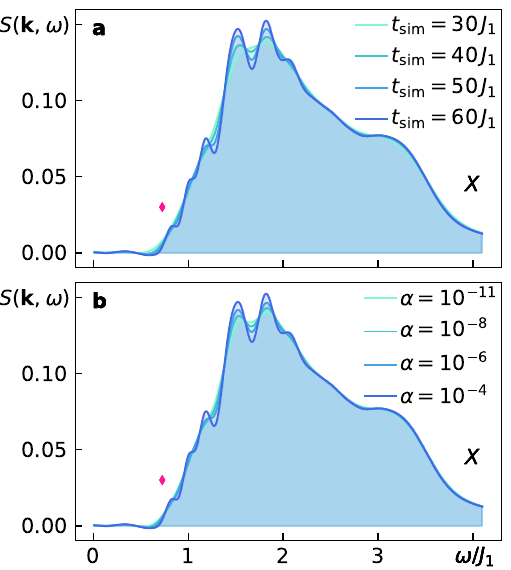}
    \caption{Convergence of the spectral function for $J_2 = 0$ at the momentum point $X$. We plot the spectral profile for various total simulation times $t_{\mathrm{sim}}$ (with $\alpha = 10^{-4}$ defining the Gaussian window) (\textbf{a}) and for various Gaussian windows at a constant simulation time of $t_{\mathrm{sim}} = 60\,J_1$ (\textbf{b}). The data has been obtained from a simulation with fixed $k_y$ at $\chi = 2000$ and cylinder length $L_x = 126$. We observe that the variation of both parameters has very similar effects on the spectral profile. $\alpha$ determines basically the value of the Gaussian envelope at the maximal evolution time (followed only by data points from linear prediction). A smaller value of $\alpha$ thus means effectively that late-time correlations are increasingly suppressed. The lowest mode that results from the level repulsion of the magnon branch and has only little spectral weight, develops as a distinct mode only for long evolution times and less suppression through the Gaussian function. The red diamond indicates the minimal excitation energy as obtained from the quasiparticle ansatz. It coincides nicely with the lowest mode that delevops in the full spectral function.}
    \label{fig:convergence_X_J20}
\end{figure}

\begin{figure}
    \centering
    \includegraphics[scale=1]{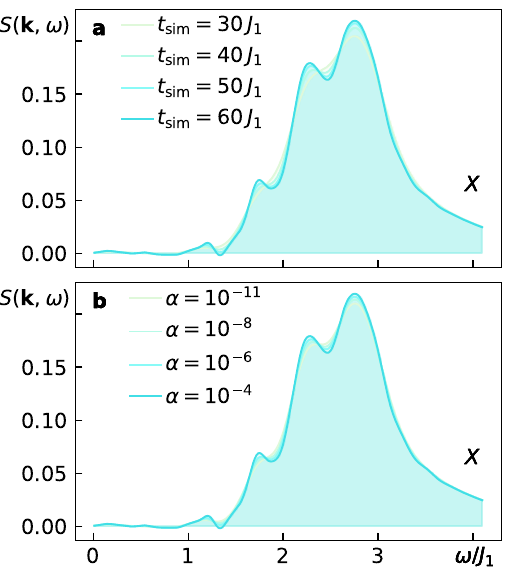}
    \caption{Spectral profiles at $X$ in the tetrahedral phase at $J_2/J_1 = 0.125$ and $J_\chi/J_1 = 0.5$. The simulation has been performed on a $\mathrm{YC}6$ system with $L_x = 81$ and $\chi = 2000$. The settings for computing the spectral function are identical to those in Fig.~\ref{fig:convergence_X_J20}. We observe a weak repelled mode in the low energy part of the spectrum (just above $\omega = 1\,J_1$, see also Fig.~{\color{blue}6} in the main text).}
    \label{fig:convergence_X_JX05}
\end{figure}

\begin{figure}
    \centering
    \includegraphics[scale=1]{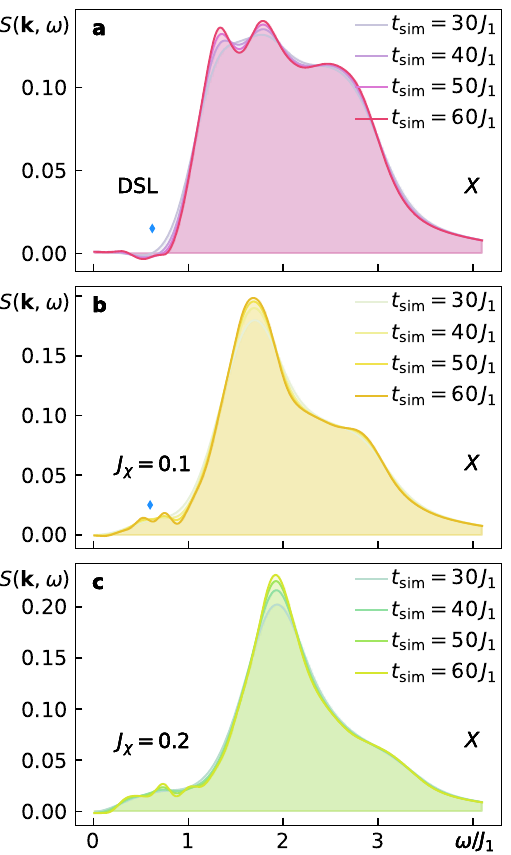}
    \caption{Convergence of the spectral profiles at $X$ with simulation time $t_{\mathrm{sim}}$. We show the data for the Dirac spin liquid at $J_2/J_1 = 0.125$ and $J_\chi = 0$ (\textbf{\color{blue}a}), and for two point in the chiral spin-liquid phase: $J_\chi/J_1 = 0.1$ (\textbf{\color{blue}b}) and $J_\chi/J_1 = 0.2$ (\textbf{\color{blue}c})---$J_2/J_1 = 0.125$ in all cases. (The Gaussian window defined as $\alpha = 10^{-6}$ as in Fig.~{\color{blue}7} of the main text.) In subplots \textbf{\color{blue}b} and \textbf{\color{blue}c}, the data was obtained from $k_y$-resolved simulations. The blue diamond indicates the lowest energy mode obtained from the MPS quasiparticle ansatz. From the DSL to the chiral spin liquid at $J_\chi/J_1 = 0.1$, the result from the latter method softens slightly. In the full spectral function, we observe an elongated tail of spectral weight below the dominant peak in contrast to the DSL phase. This can be related to the spinon-continuum as predicted by Kalmeyer and Laughlin (cf. \cite{Kalmeyer1989} and main text).}
    \label{fig:convergence_X_csl_dsl}
\end{figure}

\begin{figure}
    \centering
    \includegraphics[scale=1]{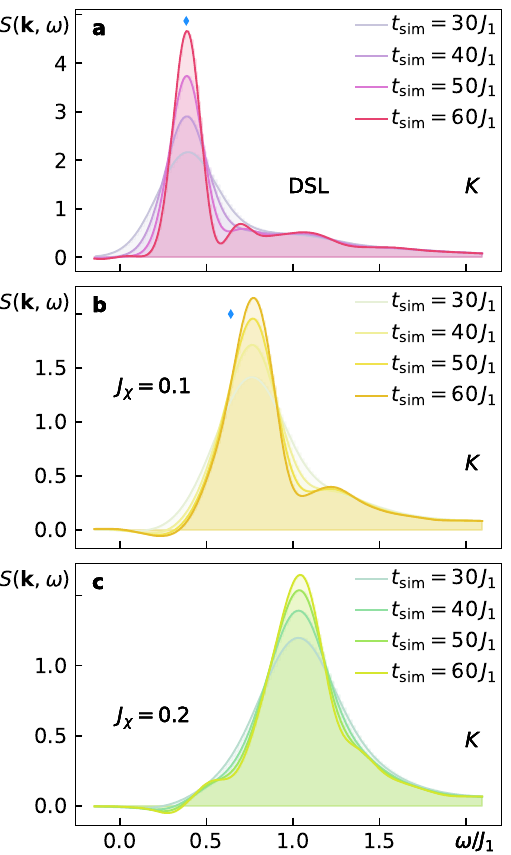}
    \caption{Spectral profiles at $K$ for the same settings as in Fig.~\ref{fig:convergence_X_csl_dsl}. In the DSL phase, we observe a sharp peak that coincides perfectly with the result from the excitation ansatz (blue diamond). The signal broadens substantially in the chiral spin liquid, where, similarly as at $X$, low-energy spectral weight emerges. In particular in \textbf{\color{blue}c} (and less pronounced in \textbf{\color{blue}b}), a kink in the profile below the maximum is indicative of a possible spinon continuum such as at the $X$ point. For $J_\chi = 0.1$, the lowest excitation energy from the quasiparticle ansatz lies below the maximum of the spectral profile, supporting the notion of relevant lower-energy excitations. The data in subplot \textbf{\color{blue}c} was obtained from a momentum-resolved time evolution.}
    \label{fig:convergence_K_csl_dsl}
\end{figure}

\begin{figure}
    \centering
    \includegraphics[scale=1]{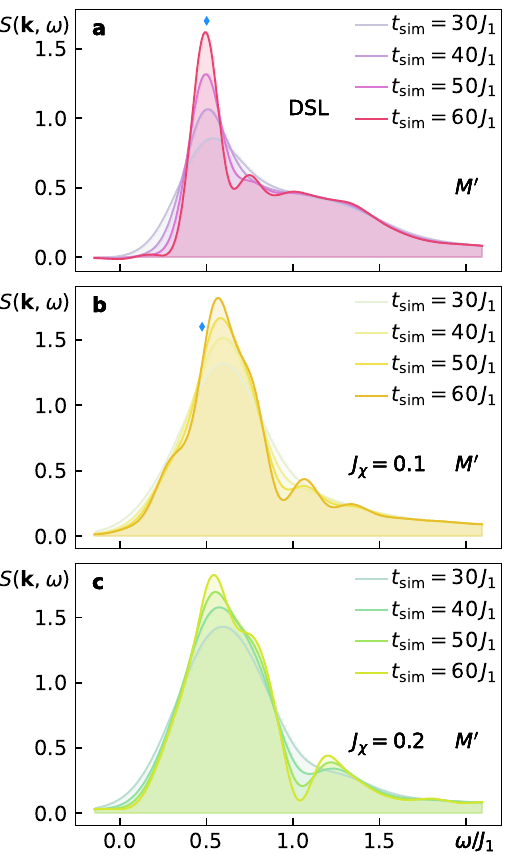}
    \caption{Spectral profiles at $M^\prime$ for the identical settings as in Fig.~\ref{fig:convergence_K_csl_dsl}. From the DSL toward the CSL phase, we observe again a broadening of the signal with the formation of low-energy spectral weight that comes along with kinks in the profile (see especially \textbf{\color{blue}b}). Whereas the quasiparticle ansatz results in remarkable agreement with the maximum of the spectral weight distribution in the DSL, the energy for $J_\chi = 0.1$ is shifted downwards from the maximum (also compared to the quasiparticle result in the DSL). This is suggestive of lower-energy modes at this momentum point.}
    \label{fig:convergence_M_csl_dsl}
\end{figure}

Figure~\ref{fig:convergence_X_J20} shows the spectral profile at $X$ at the Heisenberg point ($J_2 = J_\chi = 0$) for a momentum-resolved simulation with $k_y$ being fixed to the cut that comprises the corresponding momentum point. The upper panel \textbf{\color{blue}a} shows the convergence of the result with increasing simulation times $t_{\mathrm{sim}}$ for a fixed Gaussian window function with $\alpha = 10^{-4}$. In contrast, panel \textbf{\color{blue}b} displays the same correlation data evaluated at various $\alpha$ for a fixed total evolution time of $t_{\mathrm{sim}} = 60\,J_1$. We observe good agreement between both approaches. Applying a small value of $\alpha$ that effectively suppresses the long-time correlations is equivalent to a shorter simulation time with larger $\alpha$. The red diamond indicates the energy of the lowest excitation as obtained from the quasiparticle ansatz~\cite{Vanderstraeten2019, VanDamme2021, haegeman_variational_2012}. It agrees nicely with the lowest mode in the full spectral function that emerges as a result of level repulsion in the $120^\circ$ order. Subtle features such as the weak repelled low-energy mode in the cut $A$-$B$ evolve only for longer evolution times that allow to resolve low-energy features at sufficient precision.

Figure~\ref{fig:convergence_X_JX05} shows the equivalent data in the tetrahedrally ordered phase at $J_\chi/J_1 = 0.5$ and $J_2/J_1 = 0.125$. The convergence behaviour is identical to the discussion in the previous paragraph. Again, we identify a low-energy mode above $\omega \gtrsim 1 \,J_1$ that results from avoided quasiparticle decay in this ordered phase. The energy of the minimum of this repelled magnon branch is higher than in the $120^\circ$ ordered phase. A depiction of the full spectral function exhibiting these features can be found in Figs.~{\color{blue} 3} and {\color{blue}6} in the main text.

Figures~\ref{fig:convergence_X_csl_dsl}-\ref{fig:convergence_M_csl_dsl} illustrate the convergence of the spectral function at the momentum points $X$, $K$ and $M^\prime$ as key points of high symmetry in the Brillouin zone for the remaining phases as discussed in the main text: the candidate Dirac spin liquid at $J_2/J_1 = 0.125$ and $J_\chi = 0$ (\textbf{\color{blue}a}), the chiral spin liquid at $J_\chi/J_1 = 0.1$ (\textbf{\color{blue}b}) and $J_\chi/J_1 = 0.2$ (\textbf{\color{blue}c}) ($J_2/J_1 = 0.125$ also in the latter cases). The panels show explicitly the resulting spectral profiles for various total simulation times ($\alpha = 10^{-6}$ being fixed in accordance with Fig.~{\color{blue} 7} in the main text). At $X$, we find a steep increase of the spectral signal in the DSL with the lowest-excitation energy from the quasiparticle ansatz (blue diamond) being located just at the onset of the finite signal. In the chiral spin-liquid phase, there is an extended low-energy region with a finite distribution of spectral weight that forms a fading tail from the strong signal towards vanishing energies. This feature is compatible with the Kalmeyer-Laughlin theory of the chiral disordered phase as described by their wavefunction ansatz~\cite{Kalmeyer1989}: At the $X$ points, as well as at the $K$ and $M$ points, the two-spinon continuum is predicted to have a minimum in energy as derived from the single-spinon dispersion. Even though quantitatively the absolute energies might deviate between our numerics and the analytical approximate calculation, the qualitative features can be related between both approaches.
Accordingly, in the Dirac spin liquid, the $K$ point exhibits a sharp maximum at the energy as predicted from the quasiparticle ansatz. For $J_\chi/J_1 = 0.1$ and $J_\chi/J_1 = 0.2$, the feature broadens considerably (with the maximum spectral intensity being reduced) and there is again a low-energy regime with evanescent spectral weight below the dominant contribution that is indicative of a possible spinon-continuum (cf. for instance Fig.~\ref{fig:convergence_K_csl_dsl}\textbf{\color{blue}c}). The energy from the variational MPS excitation ansatz resides below the maximum of the spectral signal (Fig.~\ref{fig:J20_correlations_ky}\textbf{\color{blue}b}), which is supportive of the notion of relevant modes below the prevalent one.
The momentum point $M^\prime$ shows a very similar picture. We identify extended low-energy features below the maximum of the spectral distribution (cf. Fig.~\ref{fig:convergence_M_csl_dsl}\textbf{\color{blue}b} and \textbf{\color{blue}c}). One major difference between Figs.~\ref{fig:convergence_K_csl_dsl} and \ref{fig:convergence_M_csl_dsl} is that in the latter, the maximum in spectral weight is roughly constant over all three points in parameter space considered, whereas at $K$, the DSL phase comes with a more pronounced maximum in $S({\bf k}, \omega)$.

Hence, we could demonstrate the stability of key signatures of the distinct ordered and paramagnetic phases we consider in our work under simulation settings and the evaluation parameters for the spectral function.
\vfill
\end{document}